\RequirePackage{lineno} 
\documentclass[aps,prb,twocolumn,showpacs,showkeys,groupedaddress,superscriptaddress]{revtex4}  % for double-spaced preprint
\usepackage{graphicx}  % needed for figures
\usepackage{dcolumn}   % needed for some tables
\usepackage{bm}        % for math
\usepackage{pdfpages} 
\usepackage{amssymb}   % for math
\usepackage[ansinew]{inputenc}
\usepackage{hyperref}
\usepackage{rotating}
\usepackage{multirow} 
\usepackage{float}
\usepackage{color}

\begin{document}

%\setpagewiselinenumbers
%\modulolinenumbers[5]
%\linenumbers

	\title{Nonequilibrium transport in density-mo\-du\-lated phases of the second Landau level}
	
	\author{S.~Baer\footnote{Author to whom any correspondence should be addressed}}
	\author{C.~R\"ossler} 
	\author{S.~Hennel}
	\affiliation{Solid State Physics Laboratory, ETH Z\"urich, 8093 Z\"urich, Switzerland}
	\author{H.C.~Overweg} 
	\author{T.~Ihn}
	\author{K.~Ensslin}
	\author{C.~Reichl}
	\author{W.~Wegscheider} 
	\affiliation{Solid State Physics Laboratory, ETH Z\"urich, 8093 Z\"urich, Switzerland}
	\email{sbaer@phys.ethz.ch}
	
	\date{\today}
	

	\begin{abstract}
We investigate non-equilibrium transport in the reentrant integer quantum Hall phases of the second Landau level. 
At high currents, we observe a transition from the reentrant integer quantum Hall phases to classical Hall-conduction. 
Surprisingly, this transition is markedly different for the hole- and electron sides of each spin-branch. While the hole bubble phases exhibit a sharp transition to an isotropic compressible phase, the transition for the electron side occurs via an intermediate phase. 
%This behavior might be understood in terms of a current-driven two-dimensional melting transition, either taking place as a %first order 
%phase transition or as two continuous transitions involving an intermediate phase. 
This might indicate a more complex structure of the bubble phases than currently anticipated, or a breaking of the particle-hole symmetry. Such a symmetry breaking in the second Landau level might also have consequences for the physics at filling factor $\nu$=5/2.
	\end{abstract}

\maketitle
\section{Introduction}
The properties of the lowest Landau level ($N$=0) are strongly influenced by fractional quantum Hall (FQH) physics, giving rise to a large number of incompressible ground states with a vanishing longitudinal resistance $R_{xx}$ and a quantized Hall resistance $R_{xy}=h/(\nu e^2)$ at the corresponding filling factors $\nu$.
In contrast, the physics of higher Landau levels (LLs), $N\geq2$, is dominated by density-modulated quantum Hall phases. Close to half-filling, highly anisotropic and nonlinear transport properties were found \cite{lilly_anisotropic_1999,du_strongly_1999,cooper_insulating_1999}. At $\nu\approx4+1/4$ and $\nu\approx4+3/4$, $R_{xx}$ was found to vanish, while $R_{xy}$ was restored to the value of the neighboring integer quantum Hall (IQH) plateau \cite{lilly_anisotropic_1999,du_strongly_1999,cooper_insulating_1999}. This effect was referred to as reentrant integer quantum Hall (RIQH) effect. 
Theoretical \cite{fogler_ground_1996,moessner_exact_1996,shibata_ground-state_2001,koulakov_charge_1996,fogler_stripe_2001,cote_dynamics_2003} and experimental \cite{lilly_anisotropic_1999,du_strongly_1999,cooper_insulating_1999,lewis_microwave_2002,lewis_evidence_2004,lewis_microwave_2005,cooper_observation_2003} evidence points towards density modulated stripe or bubble phases which are responsible for the resistance anisotropy or the RIQH effect. 
%In the electron liquid crystal picture, quantum fluctuations lead to a modified phase diagram for the stripe phase. Depending on the strength of quantum fluctuations, the stripe phase appears as stripe crystal, smetic state, nematic state or as isotropic liquid \cite{fradkin_liquid-crystal_1999}.
In the second LL ($N$=1), a competition between FQH states and RIQH states is observed \cite{eisenstein_insulating_2002,xia_electron_2004,kumar_nonconventional_2010,deng_collective_2012,deng_nu52_2014,csathy_tilt-induced_2005,pan_experimental_2008,nuebler_density_2010}.
At certain filling factors, theory suggests the existence of two one- or two-electron-bubble and two one- or two-hole-bubble phases, which provide a lower ground-state energy than the FQH states or an isotropic liquid \cite{goerbig_microscopic_2003,goerbig_competition_2004}. 
The underlying physics of the FQH states in the second LL, like the $\nu$=5/2 and $\nu$=12/5 states is an open issue under intensive theoretical %\cite{moore_nonabelions_1991,levin_particle-hole_2007,lee_particle-hole_2007,wen_non-abelian_1991,blok_many-body_1992,morf_transition_1998,rezayi_incompressible_2000,feiguin_density_2008,moller_paired_2008,feiguin_spin_2009,zozulya_entanglement_2009,storni_fractional_2010,moller_neutral_2011,wan_edge_2006,wan_fractional_2008,dimov_spin_2008,wojs_landau-level_2010,rezayi_breaking_2011} 
and experimental %\cite{radu_quasi-particle_2008,tiemann_unraveling_2012,baer_experimental_2014-1,wurstbauer_resonant_2013,dolev_observation_2008,pan_experimental_2001,nuebler_density_2010} 
investigation \cite{willett_quantum_2013} and of great interest, due to potential applications in topological quantum computation. 
Experiments which investigate the RIQH phases of the second LL require very low temperatures and two-dimensional electron gases (2DEGs) of very high mobility. Recently temperature-dependent measurements have revealed the importance of Coulomb interactions for the formation of RIQH states and indicated that a particle-hole asymmetry in the energy scales for the formation of electron and hole bubble phases occurs \cite{deng_collective_2012,deng_nu52_2014}. A particle-hole symmetry breaking might have far-reaching consequences for the physics of the second LL, for example, for the ground-state wave function at $\nu$=5/2. A deeper understanding of the bubble phases which compete with these FQH states might be necessary to understand the physics of the second LL as a whole.

We report on magneto-transport measurements of high mobility 2DEGs. We investigate the RIQH states of the second LL in non-equilibrium transport, by driving a finite DC current bias through the system. For large DC currents, RIQH states disappear and the Hall resistance approaches its high-temperature limit, where no density modulated phases are formed. We denote this phase as the isotropic compressible phase (ICP). Surprisingly, the qualitative form of the transition from RIQH phases to the ICP is different for electron and hole bubble states. While hole bubble states exhibit a sharp transition to the ICP, a gradual transition involving a different intermediate phase is found for the electron bubble states.
Qualitatively similar findings were obtained with three different samples, made from different high mobility heterostructures which employ different growth techniques \cite{reichl_increasing_2014}. From these measurements, we extract and compare energy scales for the different RIQH states of the second LL. 
The local formation of RIQH states has been investigated by measuring transport through a quantum point contact (QPC). 
Here, signatures of the hole bubble states are completely absent, whereas signatures corresponding to the electron bubble states are most likely a pure bulk effect, while no RIQH phases are formed in the QPC channel. 
The direction dependence of the breakdown of the RIQH phases with respect to the current orientation has also been investigated. We observe that qualitative features of the transition to the isotropic compressible phase do not depend on the current orientation.

\section{Experimental details}
Measurements have been performed on photolithographically defined 500 $\mu$m wide Hall-bars, contacted with Au\-/Ge\-/Ni Ohmic contacts in a standard four-terminal measurement scheme. 
An AC current of typically $I_\mathrm{AC}\approx$ 0.5 nA is passed from source to drain and $dV_{xx}/dI_\mathrm{SD}$ and $dV_{xy}/dI_\mathrm{SD}$ are measured using lock-in techniques. In non-equilibrium situations, an additional DC current is added to the AC current. % $I_\mathrm{SD}=I_\mathrm{DC}+I_\mathrm{AC}$. 
%Currents are defined by applying a voltage across large resistors with typically $R=$ 100 M$\Omega$ - 1 G$\Omega$.
Three different high mobility structures have been used for the experiments (wafer A: $n_\mathrm{s}\approx 2.2 \times 10^{11}~\mathrm{cm}^{-2},~\mu \approx 1.9 \times 10^{7}~\mathrm{cm}^2/\mathrm{Vs}$ without illumination, wafer B: $n_\mathrm{s}\approx 2.3 \times 10^{11}~\mathrm{cm}^{-2},~\mu \approx 2.3 \times 10^{7}~\mathrm{cm}^2/\mathrm{Vs}$ without illumination, wafer C: $n_\mathrm{s}\approx 3.1 \times 10^{11}~\mathrm{cm}^{-2},~\mu \approx 1.8 \times 10^{7}~\mathrm{cm}^2/\mathrm{Vs}$ after illumination).
%Three different high mobility structures have been used for the experiments. The wafers A ($n_\mathrm{s}\approx 2.2 \times 10^{11}~\mathrm{cm}^{-2},~\mu \approx 1.9 \times 10^{7}~\mathrm{cm}^2/\mathrm{Vs}$) and B ($n_\mathrm{s}\approx 2.3 \times 10^{11}~\mathrm{cm}^{-2},~\mu \approx 2.3 \times 10^{7}~\mathrm{cm}^2/\mathrm{Vs}$) are optimized for the observation of the $\nu=5/2$ state without prior LED illumination \cite{reichl_increasing_2014}. 
%The wafer C ($n_\mathrm{s}\approx 3.1 \times 10^{11}~\mathrm{cm}^{-2},~\mu \approx 1.8 \times 10^{7}~\mathrm{cm}^2/\mathrm{Vs}$ after illumination) has been illuminated at $T\approx$ 10 K using a red LED, to allow for the observation of a fully quantized $\nu$=5/2 state and pronounced RIQH states.
Experiments have been conducted in two cryogen-free dilution refrigerators, the first with a bath temperature of approx. 10 mK and an electronic base temperature $T_\mathrm{el}$ $\approx$ (12.5 $\pm$ 0.5) mK, the second with a bath temperature of approx. 3.5 mK. The electronic temperatures have been achieved by low-pass filtering and thermally anchoring the cabling at every temperature stage (for details see Ref. \onlinecite{baer_transport_2014}). %The bath temperature ($T_\mathrm{bath}\approx10$ mK) is measured with a SQUID-based noise thermometer, which should give reliable results down to temperatures below 10 mK \cite{engert_noise_2012,engert_practical_2009}.
The filling factors of the RIQH states have been calculated from their magnetic field position relative to the center of the $\nu=5/2$ plateau in $R_{xy}$. We estimate an uncertainty in the filling factors of $\Delta \nu = \pm 0.007$ for this procedure.

\begin{figure*}
\centering
\includegraphics[width=14cm]{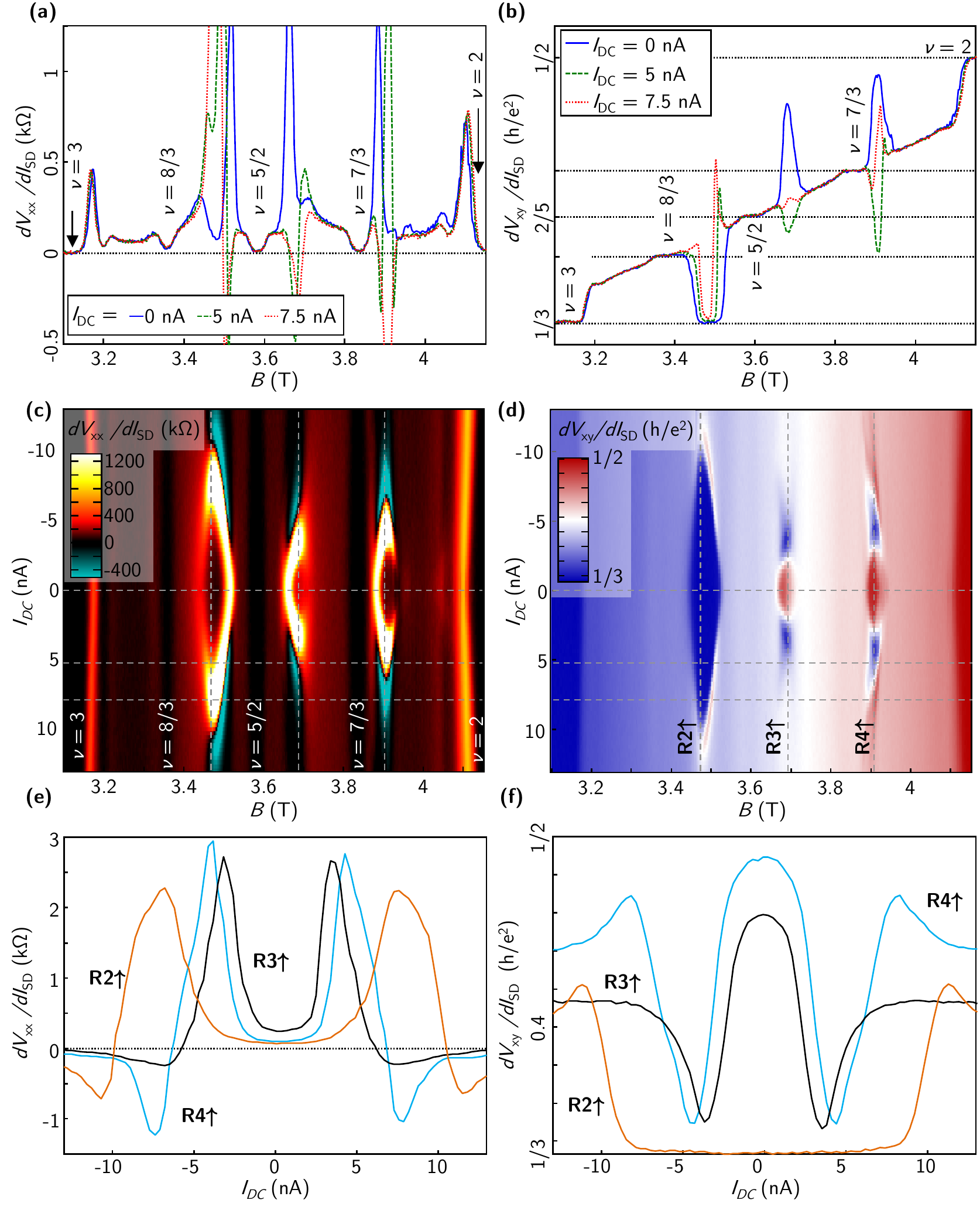}
\caption{\textbf{c:} Differential longitudinal resistance versus magnetic field and DC current bias for wafer A. Minima, associated with FQH states at $\nu=$7/3, 8/3 and 5/2 show up as black stripes. The RIQH states appear as a minimum in the longitudinal resistance with strong side-peaks and span diamond-shaped regions in the $B$-field - $I_\mathrm{DC}$- plane. At larger $I_\mathrm{DC}$ biases, negative differential resistance is found (turquoise areas). Cuts in $B$-field and bias direction are shown in \textbf{a,e}. \textbf{d}: In the differential Hall resistance, RIQH states are visible as diamond-shaped regions, where the Hall resistance tends towards the quantized resistance value of a neighboring IQH plateau. As the current bias is increased, the RIQH states which we denote as R3$\uparrow$ (black in e,f) and R4$\uparrow$ (turquoise in e,f) exhibit a strong decrease of the differential Hall resistance before they disappear in an isotropic background, corresponding to the isotropic compressible phase. In contrast, a sharp transition to the isotropic compressible phase is found for the RIQH state R2$\uparrow$ (orange in e,f).  Traces of the differential longitudinal and Hall resistances in $B$-field and in current bias direction are shown in \textbf{b,f}.}\label{RIQHE-Bias}
\end{figure*}

\section{Results and discussion}
In linear transport, only the AC current ($I_\mathrm{SD}\approx 0.5$ nA) is passed along the long axis of the Hall-bar, between source and drain and we measure the longitudinal voltage $V_{xx}$ and the Hall voltage $V_{xy}$. 
Fig. \ref{RIQHE-Bias}.a and \ref{RIQHE-Bias}.b (blue curves) show the differential longitudinal and Hall resistances, $dV_{xx}/dI_\mathrm{SD}$ and $dV_{xy}/dI_\mathrm{SD}$, for filling factors $2\leq \nu \leq 3$. The longitudinal resistance vanishes for the FQH states at $\nu$=8/3, 7/3 and 5/2, while the Hall resistance exhibits plateaus at the same magnetic fields.

Turning to non-linear transport, a DC current from source to drain is added on top of the AC current: $I_\mathrm{SD}=I_\mathrm{DC}+I_\mathrm{AC}$.
Traces of the differential longitudinal and Hall resistance for finite DC currents are also shown in Fig. \ref{RIQHE-Bias}.a and Fig. \ref{RIQHE-Bias}.b. While the FQH states or the isotropic compressible phase show a weak dependence on DC current, the differential resistance depends strongly on DC current close to the RIQH states.
Fig. \ref{RIQHE-Bias}.c shows in colorscale the differential longitudinal resistance as a function of the magnetic field $B$ and the DC current $I_\mathrm{DC}$. The FQH states at $\nu$=8/3, 7/3 and 5/2 show up as black stripes at a constant magnetic field.
When the differential Hall resistance is plotted as a function of the magnetic field $B$ and the DC current $I_\mathrm{DC}$ (\ref{RIQHE-Bias}.d), RIQH states with a quantized differential resistance of $h/(2e^2)$ or $h/(3e^2)$ show up as red or blue diamond-shaped regions. 
We denote the RIQH states in the upper spin branch ($3< \nu < 4$) by R1$\downarrow$-R4$\downarrow$ and in the lower spin branch ($2< \nu < 3$) by R1$\uparrow$-R4$\uparrow$ (see Fig. \ref{RIQHE-Bias}.d). The RIQH state R1$\uparrow$ is not observed in the measurement of Fig. \ref{RIQHE-Bias}.d, which is a first indication for the different relevant energy scales for the RIQH states. We find R1$\uparrow$ to be the most fragile state, only being observed in some of our measurements. On the high magnetic field side of $\nu$=5/2, the RIQH diamonds are neighbored in $I_\mathrm{DC}$-direction by strong side-peaks of smaller differential Hall resistance. We denote this regime as intermediate current bias phase.
The regions of a differential Hall resistance of $h/(2e^2)$ (states R3$\uparrow$ and R4$\uparrow$) or $h/(3e^2)$ (state R2$\uparrow$) are defined by the inner or outer boundaries of resistance peaks in $dV_{xx}/dI_\mathrm{SD}$ (Fig. \ref{RIQHE-Bias}.d).
%The width of the RIQH phases in $B$-field shrinks as the DC current is enhanced, until an isotropic background is found at large currents ($I_\mathrm{DC}\geq$ 12 nA). On the high magnetic field side of $\nu$=5/2, the RIQH states are neighbored by strong side-peaks of smaller differential Hall resistance. We denote this regime as intermediate current bias phase. 
%Comparing Fig. \ref{RIQHE-Bias}.b with Fig. \ref{RIQHE-Bias}.a, we see that the boundaries of the RIQH phases in the longitudinal resistance are defined by two strong side peaks, with a region of small differential longitudinal resistance in-between. As $I_\mathrm{DC}$ is increased, the side-peaks move together, resulting in a diamond shaped region. 
%The outer boundary spanned by the side peaks at $B\approx$ 3.48 T coincides with the extent of the RIQH phase at $\nu=h/(3e^2)$ in the differential Hall resistance. In contrast, the RIQH phase at $\nu=h/(2e^2)$ is defined by the inner boundary of the side peaks in $dV_{xx}/dI_\mathrm{SD}$. 
%At large currents where no RIQH phase is observed any more, regions of negative differential resistance are visible in $dV_{xx}/dI_\mathrm{SD}$ (turquoise areas in Fig. \ref{RIQHE-Bias}.a) \footnote{We remark that the ordinary resistance remains positive and that the negative differential resistance and the overall form of this measurement can be reproduced by numerically deriving a pure DC measurement (see S.I.).}, as previously observed in density-modulated phases in higher LLs \cite{gores_current-induced_2007}.

The DC current dependence of $dV_{xx}/dI_\mathrm{SD}$ and $dV_{xy}/dI_\mathrm{SD}$ is shown in Fig. \ref{RIQHE-Bias}.e and Fig. \ref{RIQHE-Bias}.f. Here, the quantity $dV_{xy}/dI_\mathrm{SD}$ shows a different behavior for the RIQH phases on both sides of the $\nu=5/2$ state. For the RIQH state R2$\uparrow$ (indicated in Fig. \ref{RIQHE-Bias}.d), $dV_{xy}/dI_\mathrm{SD}$ increases sharply close to the classical background value, with only slight overshoots when the RIQH state is destroyed. In contrast, for the RIQH state R3$\uparrow$, $dV_{xy}/dI_\mathrm{SD}$ shows very pronounced undershoots, before it reaches the classical background value. The RIQH state R4$\uparrow$ even shows a more complicated current dependence, where $dV_{xy}/dI_\mathrm{SD}$ first undershoots and then overshoots, before it reaches the classical background value.
Similar results have been obtained from measurements with other 2DEGs and subsequent cooldowns (see Fig. \ref{PhaseBoundaries}.a and S.I.). In contrast to transport in higher LLs \cite{gores_current-induced_2007}, the qualitative breakdown signatures of the different RIQH states in the second LL are independent of the crystal orientation of the sample and of the magnetic field direction. % The differential longitudinal resistance $dV_{xx}/dI_\mathrm{SD}$ shows strong peaks at the boundaries of the RIQH phases (presumably due to a small density gradient in the sample \cite{pan_low_2006,pan_quantization_2005}). 
%In the differential Hall resistance, a similar behavior as before is seen. The RIQH states at more than half filling of each spin branch of the second Landau level (R1$\downarrow$, R2$\downarrow$, R1$\uparrow$, R2$\uparrow$) show an abrupt transition to a classical Hall resistance (Fig. \ref{D120702A_2C}). On the other hand, RIQH states at less than half filling (R3$\downarrow$, R4$\downarrow$, R3$\uparrow$, R4$\uparrow$) have a less abrupt transition to a flat background, involving distinct undershoots of the differential Hall resistance.

\subsection{Phase diagram of reentrant integer quantum Hall phases}
\begin{figure*}
\centering
\includegraphics[width=14cm]{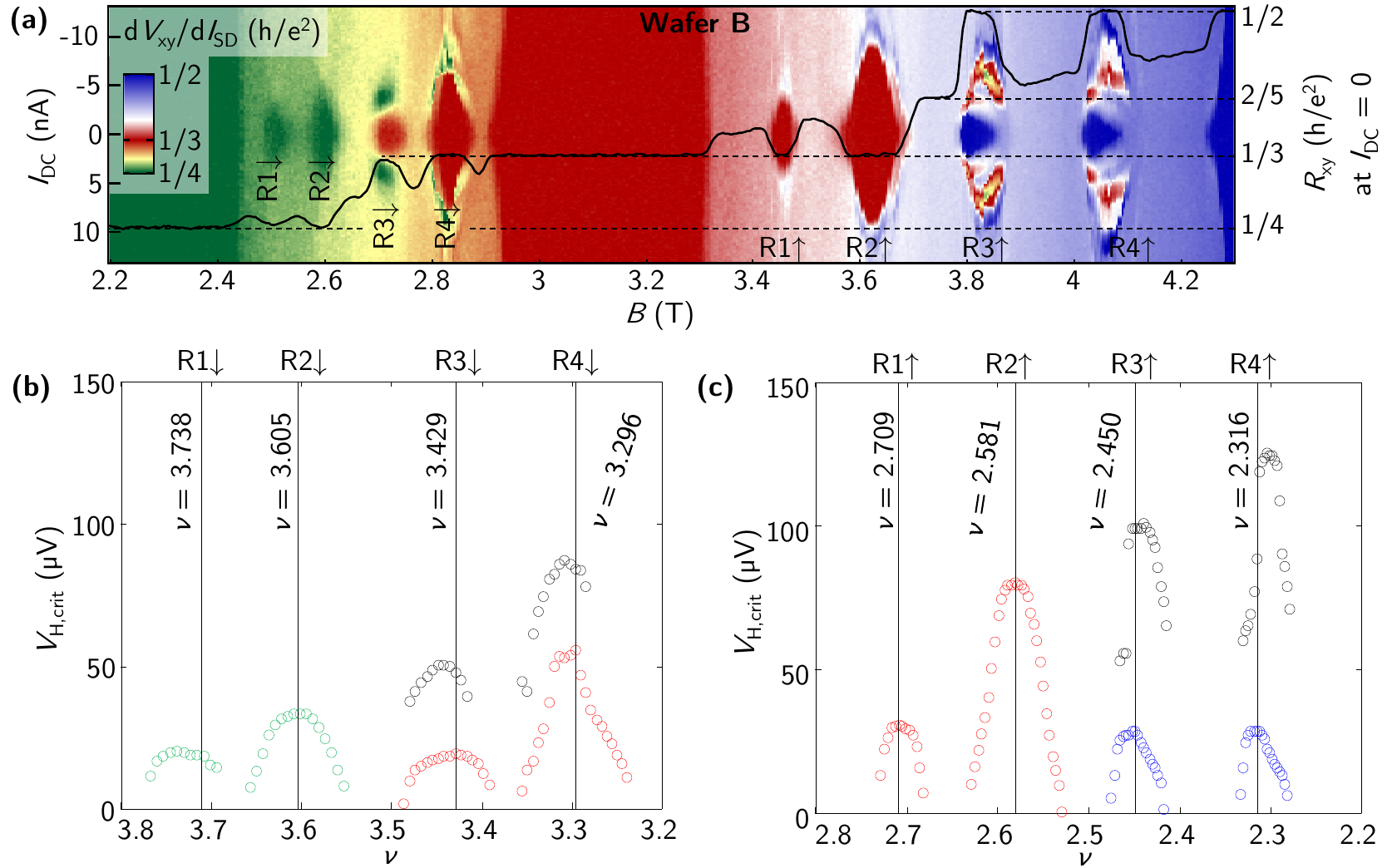}
\caption{\textbf{a}: Differential Hall resistance measured with wafer B. \textbf{b}: Critical Hall voltage for the breakdown of the RIQH states with $R_{xy}=1/4\times h/e^2$ (green) and $R_{xy}=1/3\times h/e^2$ (red) and for the transition to an isotropic background (black). \textbf{c}: Critical Hall voltage for the breakdown of the RIQH states with $R_{xy}=1/3\times h/e^2$ (red) and $R_{xy}=1/2\times h/e^2$ (blue) and for the transition to an isotropic background (black).}\label{PhaseBoundaries}
\end{figure*}

The critical current for which a RIQH state disappears is defined by the condition that the amplitude of the RIQH peak or dip in the differential Hall resistance has reached 30\% of its maximum amplitude (i.e. 30 \% of the difference between the classical Hall background resistance and the quantized resistance of the neighboring IQH plateau). The same threshold, but with different sign is used for the intermediate current bias phase of the RIQH states R3$\downarrow$, R4$\downarrow$, R3$\uparrow$ and R4$\uparrow$.
Following these considerations, we define a critical Hall voltage via $V_\mathrm{H,crit}=I_\mathrm{DC,crit}\times R_{xy, \mathrm{quant.}}$. Here, $R_{xy, \mathrm{quant.}}=h/(\nu e^2)$ is the quantized Hall resistance, corresponding to the RIQH state.
The resulting critical Hall voltage is plotted as a function of the filling factor in Fig. \ref{PhaseBoundaries} for measurement with wafer B. 

%For the RIQH states R4$\downarrow$, R3$\downarrow$, R4$\uparrow$ and R3$\uparrow$ there is a tendency that critical Hall voltages of R4$\downarrow$ and R4$\uparrow$ are larger than those of R3$\downarrow$ and R3$\uparrow$. Similarly, among the RIQH states R2$\downarrow$, R1$\downarrow$, R2$\uparrow$ and R1$\uparrow$ the largest critical Hall voltage is found for R2$\downarrow$ and R2$\uparrow$. 
The critical Hall voltage of the RIQH states R2$\downarrow$/$\uparrow$ and R4$\downarrow$/$\uparrow$ is larger than for R1$\downarrow$/$\uparrow$ and R3$\downarrow$/$\uparrow$. In contrast to the critical current, the critical Hall voltage does increase monotonically with decreasing filling factor over $\nu$=7/2 and $\nu$=5/2. The critical current exhibits a non-monotonic behavior.
A similar non-monotonic behavior has previously been observed for the critical temperatures $T_c$ at which the RIQH states start to form \cite{deng_collective_2012,deng_contrasting_2012} and gives evidence for a particle-hole asymmetry in the corresponding energy scales.

\begin{figure*}
\centering
\includegraphics[width=14cm]{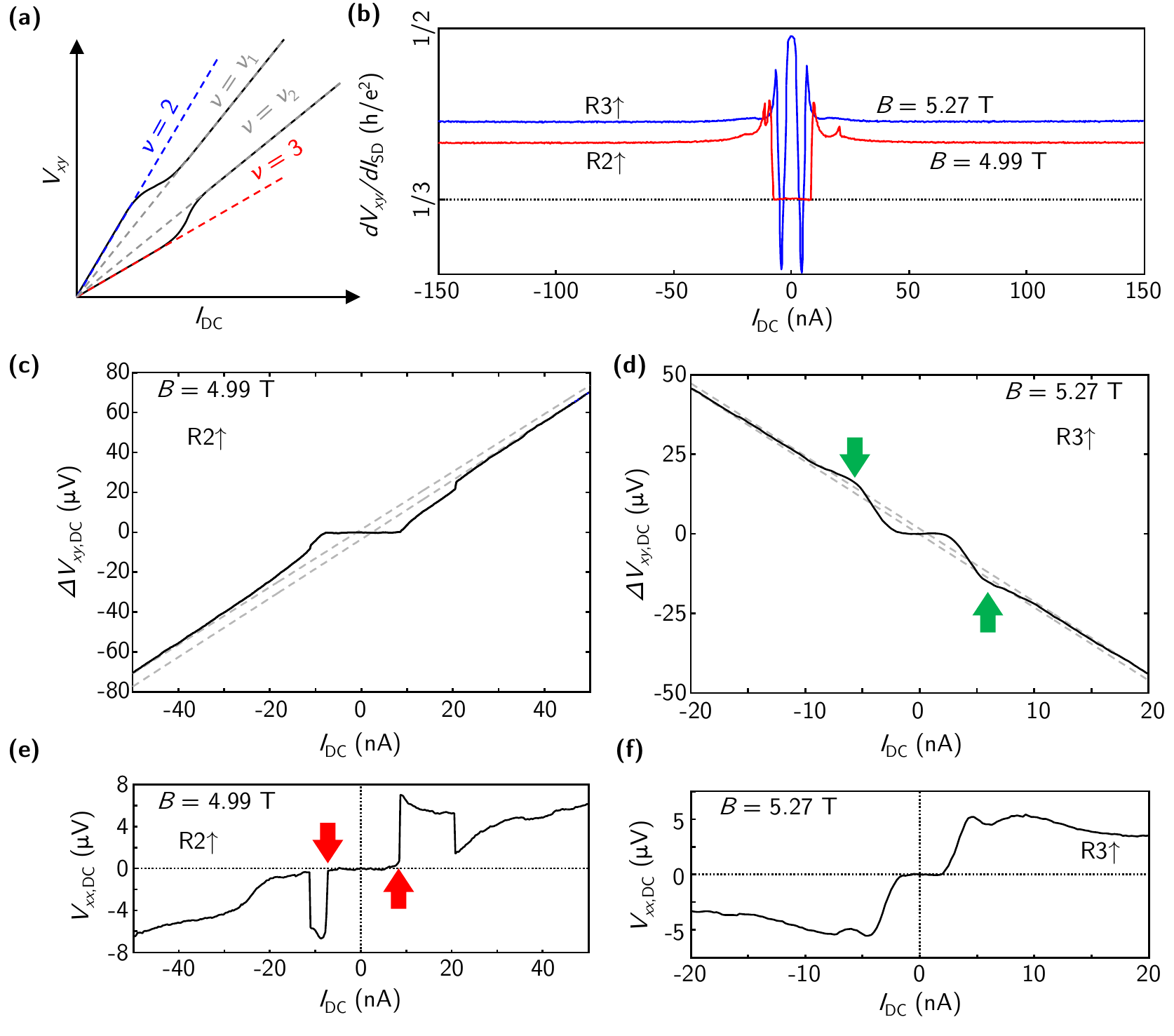}
\caption{\textbf{a:} Schematic scenario for a continuous transition to the isotropic compressible phase with non-quantized filling factors $\nu_1$ and $\nu_2$. \textbf{b:} Differential Hall resistance for the RIQH states R2$\uparrow$ and R3$\uparrow$ versus the DC current (wafer C). At $I_\mathrm{DC}>20$ nA, the differential Hall resistance is approximately constant. \textbf{c,d:} DC Hall voltage versus the DC current. Here, a linear slope corresponding to $V_{xy}=\frac{1}{3}h/e^2\times I_\mathrm{DC}$ \textbf{(c)} or $V_{xy}=\frac{1}{2}h/e^2\times I_\mathrm{DC}$ \textbf{(d)} has been subtracted. \textbf{e,f:} DC longitudinal voltage $V_{xx,\mathrm{DC}}$ versus the DC current. For the RIQH state R2$\uparrow$, jumps in the DC voltage (\textbf{(e)}, marked by red arrows) are observed.}\label{DCResults}
\end{figure*}

\subsection{Transition from RIQH phase to isotropic compressible phase}
The transition from a RIQH phase to the ICP occurs in a qualitatively different way for the RIQH states at both sides of $\nu=5/2$ and $\nu=7/2$.
The behavior of the differential Hall resistance can be better understood from a plot of the Hall voltage $V_{xy}$ versus the DC current $I_\mathrm{DC}$, $V_{xy}(I_\mathrm{DC})$.
We schematically show the expected dependence of $V_{xy}$ on $I_\mathrm{DC}$ in Fig. \ref{DCResults}.a for a continuous transition between a RIQH phase and an isotropic compressible phase. For a constant filling factor, a linear slope $V_{xy}=h/(\nu e^2) \times I_\mathrm{DC}$ is expected.
Hence at a small DC current, where the RIQH state exists, the RIQH states R3$\uparrow$ and R4$\uparrow$ (with a plateau value of $h/2e^2$) exhibit a slope of $h/2e^2$ in the $I_\mathrm{DC}$-$V_{xy}$ diagram (green dashed line in Fig. \ref{DCResults}.a). Similarly, a slope of $h/3e^2$ is expected for the RIQH states R1$\uparrow$ and R2$\uparrow$ (red dashed line in Fig. \ref{DCResults}.a). 
In the limit of large currents, the RIQH phase does no longer exist, and a background Hall resistance, which is determined by the corresponding filling factor, is found (grey dashed lines in Fig. \ref{DCResults}.a). 
Thus, the curve $V_{xy}(I_\mathrm{DC})$ is expected to possess a slope of $h/\nu e^2$ for large DC currents, where $\nu\approx$ 2.43 or 2.30 for the RIQH states R3$\uparrow$ and R4$\uparrow$ and  $\nu\approx$ 2.71 or 2.58 for R1$\uparrow$ and R2$\uparrow$. 
In-between the low- and high-current regime, a continuous transition occurs, while all the linear slopes discussed before are expected to interpolate to the origin of the diagram. In order to satisfy these conditions, an intermediate regime with a smaller slope of $V_{xy}(I_\mathrm{DC})$ (hence smaller differential Hall resistance) is expected for the transition from $\nu=2$ to the isotropic liquid. In contrast, for the transition from $\nu$=3, we expect an intermediate regime with a larger slope in $V_{xy}$ versus $I_\mathrm{DC}$.
In the differential resistance, this would be visible as pronounced undershoots of the differential resistance for the RIQH states R3$\uparrow$ and R4$\uparrow$, while strong overshoots should be visible for R1$\uparrow$ and R2$\uparrow$.
The expected $V_{xy}(I_\mathrm{DC})$ for such a continuous transition are shown schematically as black lines in Fig. \ref{DCResults}.a, where $\nu_1$ and $\nu_2$ denote the non-quantized ``background" filling factors.

The measured transition from RIQH phases to the ICP can be compared to this expectation. A measured differential Hall resistance is plotted versus the DC current in Fig. \ref{DCResults}.b. As the DC current is increased at the magnetic field values corresponding to the RIQH states R2$\uparrow$ and R3$\uparrow$, a transition to an isotropic compressible background occurs for DC currents smaller than $20$ nA and the differential Hall resistance is constant at higher currents (Fig. \ref{DCResults}.b).
The measured DC Hall and longitudinal voltages, $V_{xy,\mathrm{DC}}$ and  $V_{xx,\mathrm{DC}}$ are shown in Fig. \ref{DCResults}.c-f. For better visibility, a linear slope corresponding to $V_{xy}=\frac{1}{3}h/e^2\times I_\mathrm{DC}$ or $V_{xy}=\frac{1}{2}h/e^2\times I_\mathrm{DC}$ has been subtracted from the DC Hall voltage. After this subtraction, the RIQH phases are seen as plateaus at $\Delta V_{xy,\mathrm{DC}}=0$. 
Dashed lines are fitted to the linear parts of the Hall voltage (for 75 nA $<$  $\vert I_\mathrm{DC}\vert$ $<$ 150 nA).
%In contrast to before, slopes interpolate approximately through zero for all RIQH states (within the fit uncertainty).
For the RIQH state R2$\uparrow$, the DC Hall voltage abruptly deviates from the quantized slope, when increasing the DC current above a critical value of $\approx$ 10 nA (Fig. \ref{DCResults}.c). At the same time, sharp jumps are observed in the longitudinal DC voltage $V_{xx,\mathrm{DC}}$ (marked by red arrow in Fig. \ref{DCResults}.e). Above the critical current, $\Delta V_{xy,\mathrm{DC}}$ in Fig. \ref{DCResults}.c increases monotonically until it reaches the slope corresponding to the ICP, similar to the current-voltage characteristics of a gapped superconductor. In contrast, $\Delta V_{xy,\mathrm{DC}}$ changes smoothly for the RIQH state R3$\uparrow$ and jumps in $V_{xx,\mathrm{DC}}$ are absent. As the DC current is increased, overshoots below or above the linear regime of $\Delta V_{xy,\mathrm{DC}}$ (marked by green arrows in Fig. \ref{DCResults}.d) are observed. Similar observations have been made in several samples.
While our observations for the RIQH states R3$\uparrow$ and R4$\uparrow$ (and R3$\downarrow$, R4$\downarrow$) resemble the scenario for a continuous transition to the ICP as depicted in Fig. \ref{DCResults}.a, the sharp transition for the RIQH states R1$\uparrow$ and R2$\uparrow$ (and R1$\downarrow$, R2$\downarrow$) might be interpreted as a depinning of the hole bubbles, or a mechanism involving a jump of the Fermi energy across an energy gap. 
These two different observations might also be explained by different routes for a two-dimensional melting transition, either by a first-order phase transition or by two continuous transitions via an intermediate phase \cite{kosterlitz_ordering_1973,kosterlitz_critical_1974,halperin_theory_1978,young_melting_1979,fradkin_liquid-crystal_1999}.
Measurements of wafer A in a van der Pauw geometry at a bath temperature of approx. 3.5 mK indicate the formation of an intermediate current bias phase, for both electron and hole bubbles (see section \ref{OrientDep}). However, also here the intermediate current bias phase is more pronounced for the RIQH states R3$\uparrow$ and R4$\uparrow$. This might indicate that the different behavior of electron and hole bubble phases is governed by different energy scales.

\subsection{Reentrant integer quantum Hall phases in a QPC}
\begin{figure*}[h!]
\centering
\includegraphics[width=14cm]{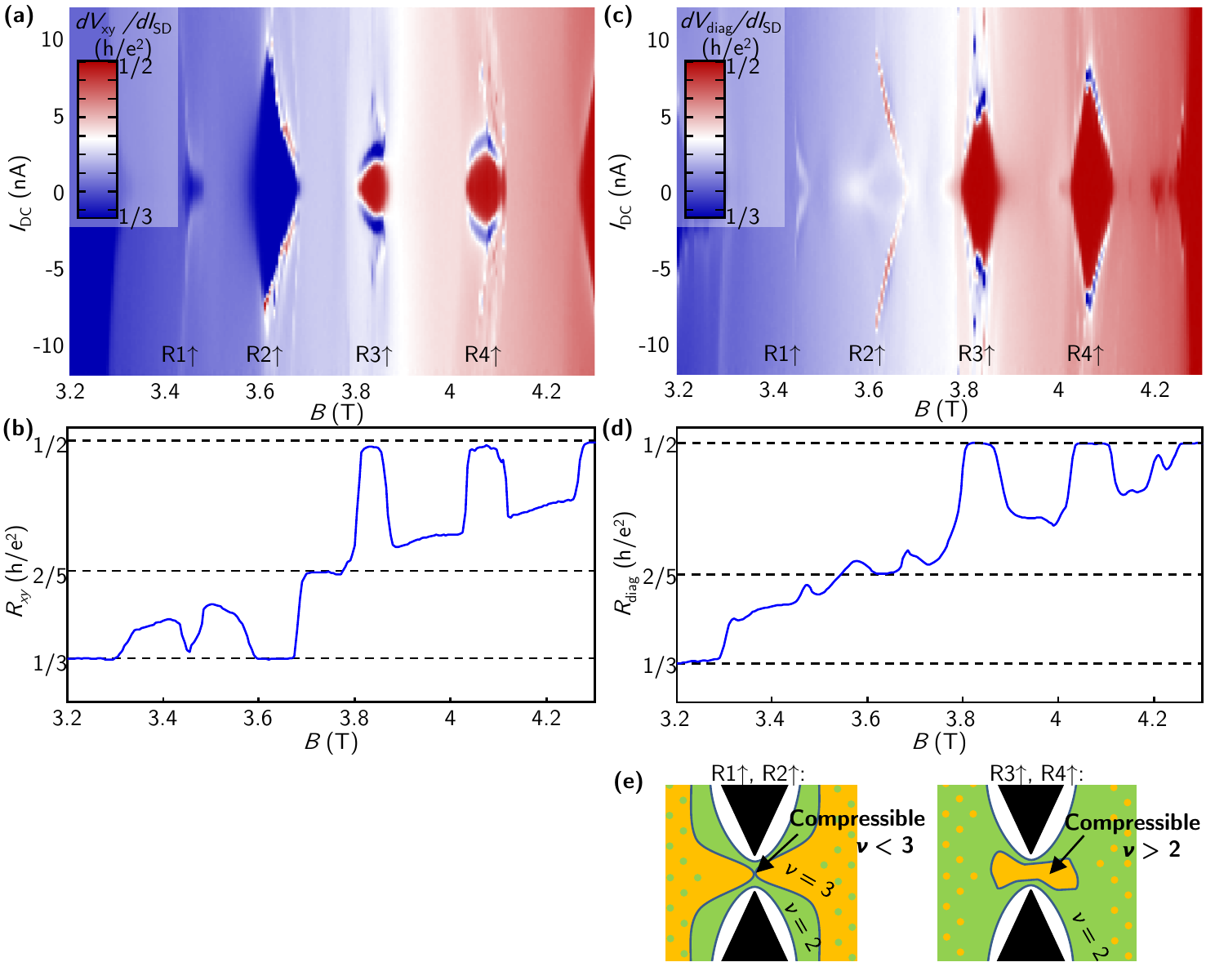}
\caption{{\textbf{a:} Differential bulk Hall resistance as a function of the DC current. \textbf{b:} Differential Hall resistance at $I_\mathrm{DC}=0$. The RIQH states R1$\uparrow$ - R4$\uparrow$ are clearly visible. \textbf{c:} Differential diagonal resistance across a 1.1 $\mu$m wide QPC. Here, - 2.3 V are applied to the top-gates. At zero DC current, the RIQH states R1$\uparrow$ and R2$\uparrow$ are no longer visible \textbf{(d)}, whereas R3$\uparrow$ and R4$\uparrow$ are even more pronounced than in the bulk. \textbf{e:} Schematic density distribution in the QPC and the bulk for the different RIQH states.}}\label{QPC-RIQHE}
\end{figure*}

More information about the local formation of RIQH states can be obtained from investigating a constriction defined by a QPC. Fig. \ref{QPC-RIQHE}.a,b shows the differential Hall resistance, measured in the bulk of a high mobility Hall-bar. As before, R1$\uparrow$ and R2$\uparrow$ show a sharp threshold to an isotropic compressible phase, while a transition via an intermediate phase is observed for R3$\uparrow$ and R4$\uparrow$.
Now we turn to a situation, where a 1.1 $\mu$m wide QPC, fabricated on top of the 2DEG is defined by applying negative voltages to the top-gates ($V_\mathrm{QPC}=-2.3$ V). We measure the voltage $V_\mathrm{diag}$ diagonally across the QPC, from which an effective filling factor of the QPC, $\nu_\mathrm{QPC}$, is obtained via $R_\mathrm{diag}=\frac{h}{e^2\nu_\mathrm{QPC}}$ \cite{beenakker_quantum_2004}.
The system is tuned to a weak backscattering situation, where bulk and QPC densities are identical, which can be seen from the overlap of the IQH plateaus in $R_{xy}$ and $R_\mathrm{diag}$ (see Ref. \onlinecite{baer_experimental_2014}). 
Figs. \ref{QPC-RIQHE}.c,d show the differential diagonal resistance for this situation. While the IQH plateaus are still clearly visible (Fig. \ref{QPC-RIQHE}.d), FQH states can no longer be clearly identified. The reason for this is weak quasiparticle tunneling which destroys the FQH plateaus and gives rise to a power-law tunneling conductance between counterpropagating edge states (see Ref. \onlinecite{baer_experimental_2014}). In this situation, the RIQH states R1$\uparrow$ and R2$\uparrow$ have vanished and are no longer seen in the diagonal resistance. In contrast, R3$\uparrow$ and R4$\uparrow$ are even more pronounced than in the bulk. The regions corresponding to a quantization at $h/(2e^2)$ have a larger extent in the $I_\mathrm{DC}$ direction than in the bulk.

The observation of the RIQH states R3$\uparrow$ and R4$\uparrow$ in the diagonal resistance however does not imply that they are formed in the QPC constriction. The relation $R_\mathrm{diag}=\frac{h}{e^2\nu_\mathrm{QPC}}$ is only valid, if $\nu_\mathrm{QPC}\leq\nu_\mathrm{bulk}$, which is not necessarily the case for the density modulated bubble phases.
In this case, the following relation holds: $R_\mathrm{diag}=\frac{h}{e^2\mathrm{min}(\nu_\mathrm{QPC},\nu_\mathrm{bulk})}$. A schematic picture of this situation is shown in Fig. \ref{QPC-RIQHE}.e. 
For the RIQH states R1$\uparrow$ and R2$\uparrow$, the innermost edge state (corresponding to $\nu$=3) is partially backscattered at the QPC and hence the quantization is lost, even if the RIQH states persist in the bulk. For the RIQH states R3$\uparrow$ and R4$\uparrow$ however, the innermost edge state of the bulk corresponds to $\nu=2$. 
If a compressible region is formed in the QPC, it has a larger filling factor than two, and hence the $\nu$=2 edge state is protected from backscattering. Hence, the quantized value of $dV_\mathrm{diag}/dI_\mathrm{DC}=h/(2e^2)$ can be interpreted as a bulk signature, where $\nu_\mathrm{bulk}=2$ describes the bulk transport properties and where a compressible phase is formed in the QPC, leading to $\nu_\mathrm{QPC}> 2$. 

We conclude that most likely no density modulated phases persist in the QPC channel. The different characteristics of the RIQH states R3$\uparrow$ and R4$\uparrow$ when measured in the bulk and across a QPC might result from a density gradient close to the QPC or from a variation in the current distribution and hence a partial local Hall voltage drop in the center of the QPC channel.

\subsection{Orientation dependence of transport in the RIQH regime}\label{OrientDep}
\begin{figure*}[h]
\includegraphics[width=14cm]{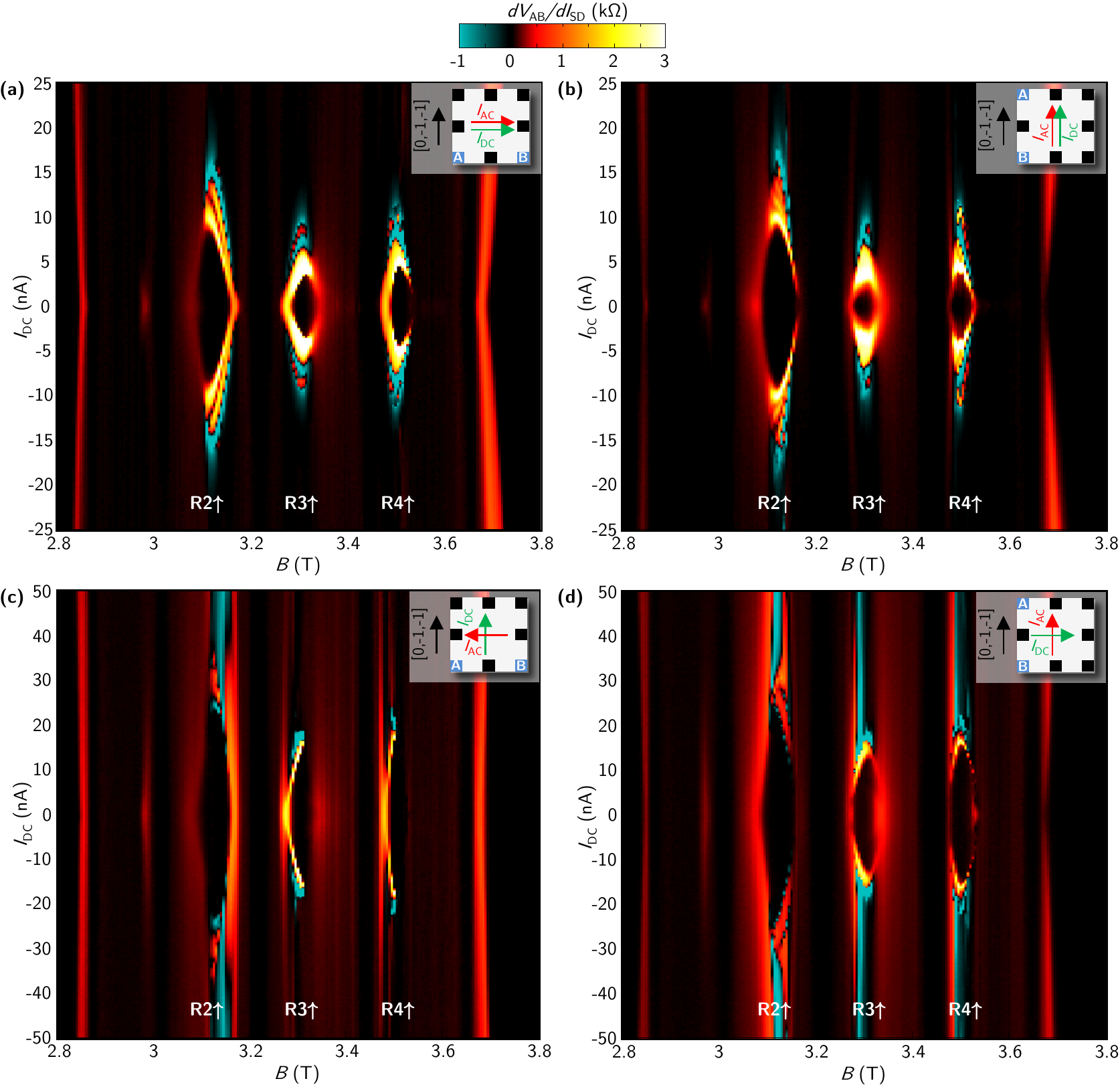}
\caption{Differential longitudinal resistance for different directions of DC and AC currents. The respective contact configurations and current directions relative to the crystal orientation are shown schematically as insets.  Measurements where AC and DC currents flow in the $x$-direction \textbf{(a)} or the $y$-direction \textbf{(b)} show similar energy scales for the transition to an isotropic background.}\label{SzymonXX}
\end{figure*}

\begin{figure*}[h]
\includegraphics[width=14cm]{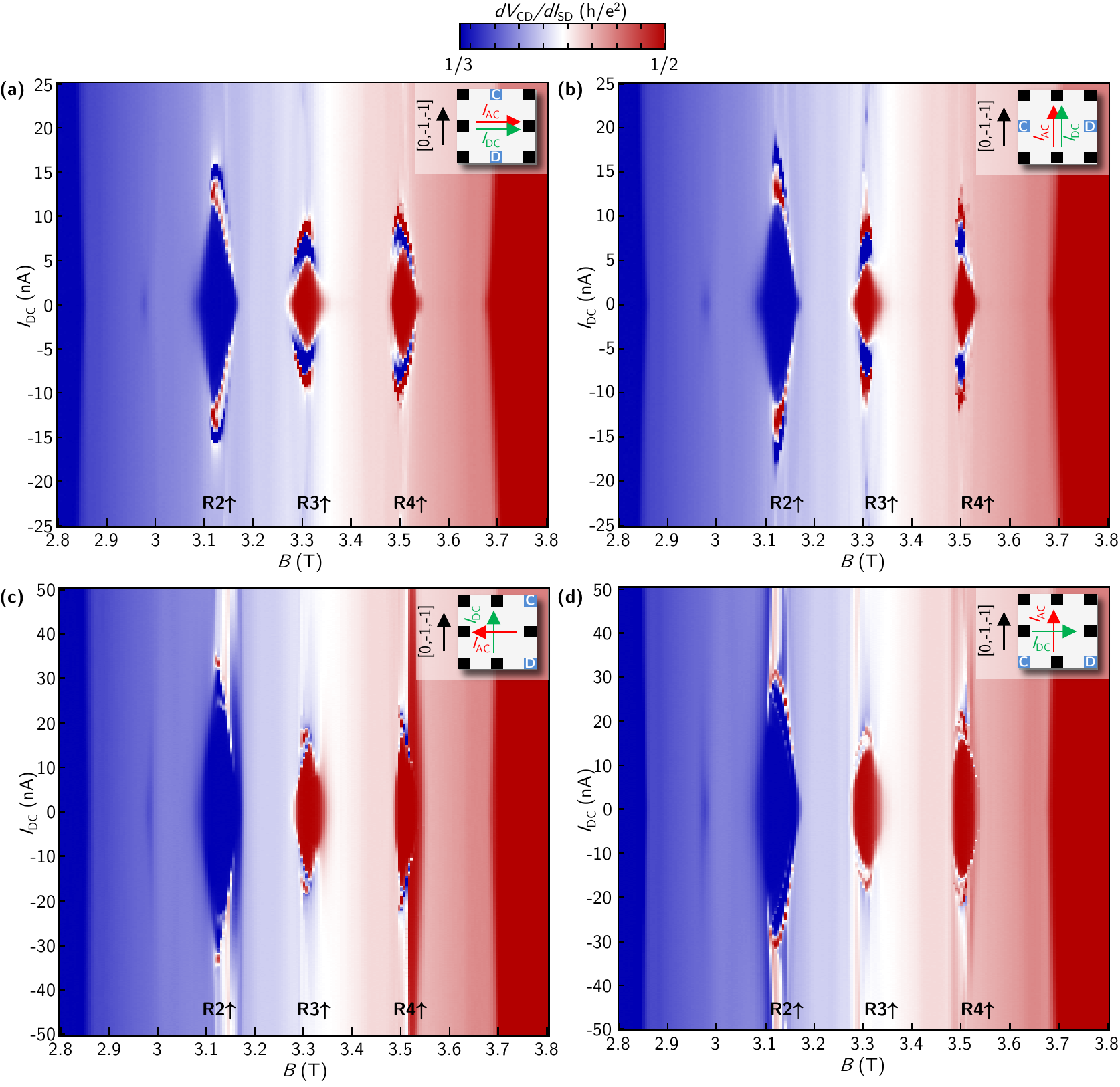}
\caption{Differential Hall resistance for different directions of DC and AC currents. The respective contact configurations and current directions relative to the crystal orientation are shown schematically as insets.  Measurements where AC and DC currents flow in the $x$-direction \textbf{(a)} or the $y$-direction \textbf{(b)} show similar energy scales for the transition to an isotropic background.}\label{SzymonXY}
\end{figure*}

To investigate the orientation dependence of the transport in the RIQH regime, we have measured a square van der Pauw geometry of wafer A with eight Au/Ge/Ni Ohmic contacts (see schematic inset Figs. \ref{SzymonXX} and \ref{SzymonXY}). The 750 $\mu$m $\times$ 750 $\mu$m square van der Pauw mesas and contacts have been defined by optical lithography, to ensure a perfect square geometry and alignment with the crystal directions.

\begin{figure}[h]
\includegraphics[width=8.5cm]{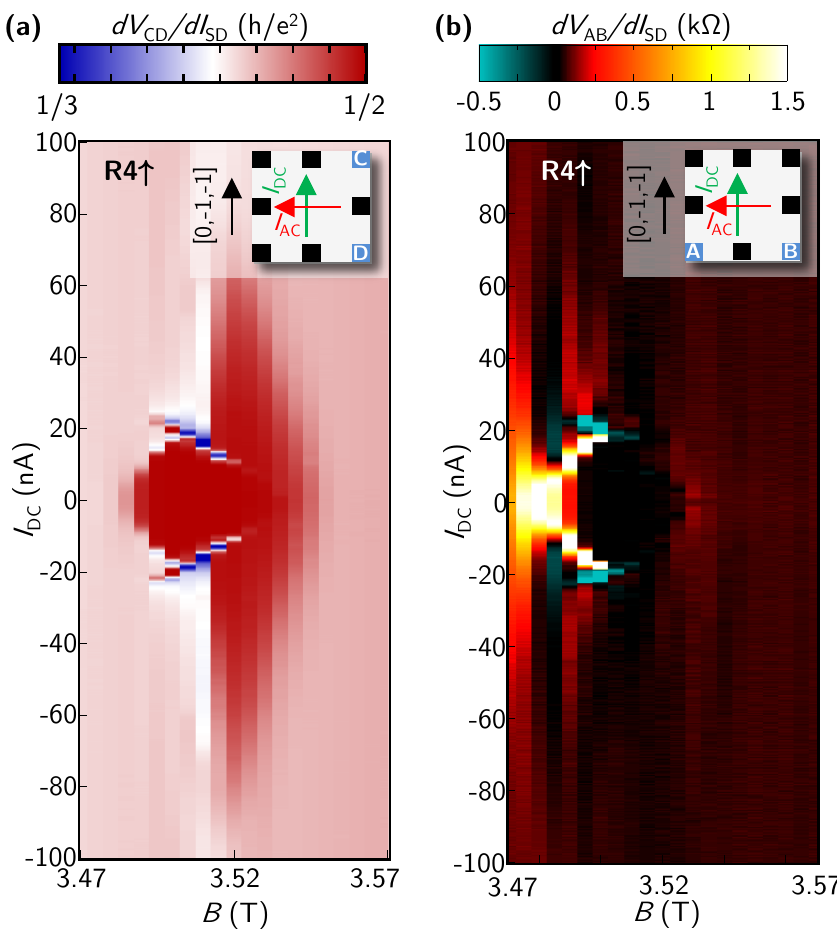}
\caption{Differential Hall (a) and longitudinal resistance (b) for the RIQH state R4$\uparrow$ and the current orientation depicted schematically in the inset.}\label{SzymonDD}
\end{figure}

In contrast to stripe phases in higher LLs, we expect the bubble phases of the second LL to be isotropic. However, it was shown that a large DC current can induce anisotropies in density modulated phases which are isotropic at small currents \cite{gores_current-induced_2007}.
Figs. \ref{SzymonXX} and \ref{SzymonXY} show the differential longitudinal and Hall resistances for different orientations of the AC and DC current as a function of magnetic field and DC current. Insets show schematically the current configurations and the contact configuration used for each measurement. For parallel AC and DC currents, breakdown signatures of the RIQH states of the second LL are qualitatively similar (Fig. \ref{SzymonXY}.a,b) and critical currents are of comparable magnitude. %When AC and DC current are passed either along the long axis of the Hall-bar ($x$-direction, Fig. \ref{SzymonXY}.a) or perpendicular to the long axis ($y$-direction, Fig. \ref{SzymonXY}.b), critical currents are of comparable magnitude. 
%As before, R3$\uparrow$ and R4$\uparrow$ exhibit a transition to an isotropic phase with strong undershoots in the differential Hall resistance, while only weak overshoots are observed for R1$\uparrow$ and R2$\uparrow$.
%Qualitatively, the breakdown for both current orientations looks similar, while details like the critical Hall voltage of the RIQH phases change slightly.
However, driving the DC current perpendicular to the AC current changes the shape of the RIQH phases. The most distinct change is observed for the RIQH state R4$\uparrow$ (see Figs. \ref{SzymonXX} and \ref{SzymonXY} and a close-up in Fig. \ref{SzymonDD}). Here, the original diamond-shaped RIQH phase extends at the high magnetic field side into a larger diamond-shaped region, with a differential Hall resistance close to $\frac{h}{2e^2}$ (see Fig. \ref{SzymonDD}.a), which prevails up to large currents of approximately 90 nA. This region is not observed for transport in perpendicular direction (Fig.  \ref{SzymonXX}.d and \ref{SzymonXY}.d) and thus might arise due to current-induced anisotropies.

%
%\begin{figure*}
%\centering
%\includegraphics[width=14cm]{Szymon1.pdf}
%\caption{DC Hall voltage versus the DC current, measured in the van der Pauw geometry using wafer A. A linear slope corresponding to $V_{xy}=\frac{1}{3}h/e^2\times I_\mathrm{DC}$ or $V_{xy}=\frac{1}{2}h/e^2\times I_\mathrm{DC}$ has been subtracted.}\label{Szymon1}
%\end{figure*}

\section{Conclusion}
Our observations suggest that either the particle-hole symmetry is broken for the density modulated phases in the second Landau level, or that the bubble phases are of a more complex structure than currently anticipated. A symmetry breaking has a strong influence on the properties of the density modulated phases and might also have an important impact on the physics of the FQH states in the second LL, like the $\nu$=5/2 FQH state.

We gratefully acknowledge discussions with \mbox{Bernd Rosenow}, \mbox{Jurgen Smet}, \mbox{Rudolf H. Morf}, \mbox{Lars Tiemann} and \mbox{Werner Dietsche}. 
We acknowledge the support of the ETH FIRST laboratory and financial support of the
Swiss National Science Foundation (Schweizerischer Nationalfonds, NCCR ``Quantum Science and Technology").
H.C. O. 	acknowledges financial support of the ICFP, D\'epartement de Physique, Ecole Normale Sup\'erieure, 75005 Paris, France.

Note added: during the preparation of the manuscript, a preprint by Rossokhaty et al. appeared, which discusses the high-current breakdown of the RIQH phases \cite{rossokhaty_bias-induced_2014}.
\section{References}
\bibliographystyle{apsrev4-1}
\bibliography{Bibliothek}

\begin{thebibliography}{36}%
\makeatletter
\providecommand \@ifxundefined [1]{%
 \@ifx{#1\undefined}
}%
\providecommand \@ifnum [1]{%
 \ifnum #1\expandafter \@firstoftwo
 \else \expandafter \@secondoftwo
 \fi
}%
\providecommand \@ifx [1]{%
 \ifx #1\expandafter \@firstoftwo
 \else \expandafter \@secondoftwo
 \fi
}%
\providecommand \natexlab [1]{#1}%
\providecommand \enquote  [1]{``#1''}%
\providecommand \bibnamefont  [1]{#1}%
\providecommand \bibfnamefont [1]{#1}%
\providecommand \citenamefont [1]{#1}%
\providecommand \href@noop [0]{\@secondoftwo}%
\providecommand \href [0]{\begingroup \@sanitize@url \@href}%
\providecommand \@href[1]{\@@startlink{#1}\@@href}%
\providecommand \@@href[1]{\endgroup#1\@@endlink}%
\providecommand \@sanitize@url [0]{\catcode `\\12\catcode `\$12\catcode
  `\&12\catcode `\#12\catcode `\^12\catcode `\_12\catcode `\%12\relax}%
\providecommand \@@startlink[1]{}%
\providecommand \@@endlink[0]{}%
\providecommand \url  [0]{\begingroup\@sanitize@url \@url }%
\providecommand \@url [1]{\endgroup\@href {#1}{\urlprefix }}%
\providecommand \urlprefix  [0]{URL }%
\providecommand \Eprint [0]{\href }%
\providecommand \doibase [0]{http://dx.doi.org/}%
\providecommand \selectlanguage [0]{\@gobble}%
\providecommand \bibinfo  [0]{\@secondoftwo}%
\providecommand \bibfield  [0]{\@secondoftwo}%
\providecommand \translation [1]{[#1]}%
\providecommand \BibitemOpen [0]{}%
\providecommand \bibitemStop [0]{}%
\providecommand \bibitemNoStop [0]{.\EOS\space}%
\providecommand \EOS [0]{\spacefactor3000\relax}%
\providecommand \BibitemShut  [1]{\csname bibitem#1\endcsname}%
\let\auto@bib@innerbib\@empty
%</preamble>
\bibitem [{\citenamefont {Lilly}\ \emph {et~al.}(1999)\citenamefont {Lilly},
  \citenamefont {Cooper}, \citenamefont {Eisenstein}, \citenamefont
  {Pfeiffer},\ and\ \citenamefont {West}}]{lilly_anisotropic_1999}%
  \BibitemOpen
  \bibfield  {author} {\bibinfo {author} {\bibfnamefont {M.~P.}\ \bibnamefont
  {Lilly}}, \bibinfo {author} {\bibfnamefont {K.~B.}\ \bibnamefont {Cooper}},
  \bibinfo {author} {\bibfnamefont {J.~P.}\ \bibnamefont {Eisenstein}},
  \bibinfo {author} {\bibfnamefont {L.~N.}\ \bibnamefont {Pfeiffer}}, \ and\
  \bibinfo {author} {\bibfnamefont {K.~W.}\ \bibnamefont {West}},\ }\href
  {\doibase 10.1103/PhysRevLett.83.824} {\bibfield  {journal} {\bibinfo
  {journal} {Physical Review Letters}\ }\textbf {\bibinfo {volume} {83}},\
  \bibinfo {pages} {824} (\bibinfo {year} {1999})}\BibitemShut {NoStop}%
\bibitem [{\citenamefont {Du}\ \emph {et~al.}(1999)\citenamefont {Du},
  \citenamefont {Tsui}, \citenamefont {Stormer}, \citenamefont {Pfeiffer},
  \citenamefont {Baldwin},\ and\ \citenamefont {West}}]{du_strongly_1999}%
  \BibitemOpen
  \bibfield  {author} {\bibinfo {author} {\bibfnamefont {R.~R.}\ \bibnamefont
  {Du}}, \bibinfo {author} {\bibfnamefont {D.~C.}\ \bibnamefont {Tsui}},
  \bibinfo {author} {\bibfnamefont {H.~L.}\ \bibnamefont {Stormer}}, \bibinfo
  {author} {\bibfnamefont {L.~N.}\ \bibnamefont {Pfeiffer}}, \bibinfo {author}
  {\bibfnamefont {K.~W.}\ \bibnamefont {Baldwin}}, \ and\ \bibinfo {author}
  {\bibfnamefont {K.~W.}\ \bibnamefont {West}},\ }\href {\doibase
  10.1016/S0038-1098(98)00578-X} {\bibfield  {journal} {\bibinfo  {journal}
  {Solid State Communications}\ }\textbf {\bibinfo {volume} {109}},\ \bibinfo
  {pages} {389} (\bibinfo {year} {1999})}\BibitemShut {NoStop}%
\bibitem [{\citenamefont {Cooper}\ \emph {et~al.}(1999)\citenamefont {Cooper},
  \citenamefont {Lilly}, \citenamefont {Eisenstein}, \citenamefont {Pfeiffer},\
  and\ \citenamefont {West}}]{cooper_insulating_1999}%
  \BibitemOpen
  \bibfield  {author} {\bibinfo {author} {\bibfnamefont {K.~B.}\ \bibnamefont
  {Cooper}}, \bibinfo {author} {\bibfnamefont {M.~P.}\ \bibnamefont {Lilly}},
  \bibinfo {author} {\bibfnamefont {J.~P.}\ \bibnamefont {Eisenstein}},
  \bibinfo {author} {\bibfnamefont {L.~N.}\ \bibnamefont {Pfeiffer}}, \ and\
  \bibinfo {author} {\bibfnamefont {K.~W.}\ \bibnamefont {West}},\ }\href
  {\doibase 10.1103/PhysRevB.60.R11285} {\bibfield  {journal} {\bibinfo
  {journal} {Physical Review B}\ }\textbf {\bibinfo {volume} {60}},\ \bibinfo
  {pages} {R11285} (\bibinfo {year} {1999})}\BibitemShut {NoStop}%
\bibitem [{\citenamefont {Fogler}\ \emph {et~al.}(1996)\citenamefont {Fogler},
  \citenamefont {Koulakov},\ and\ \citenamefont
  {Shklovskii}}]{fogler_ground_1996}%
  \BibitemOpen
  \bibfield  {author} {\bibinfo {author} {\bibfnamefont {M.~M.}\ \bibnamefont
  {Fogler}}, \bibinfo {author} {\bibfnamefont {A.~A.}\ \bibnamefont
  {Koulakov}}, \ and\ \bibinfo {author} {\bibfnamefont {B.~I.}\ \bibnamefont
  {Shklovskii}},\ }\href {\doibase 10.1103/PhysRevB.54.1853} {\bibfield
  {journal} {\bibinfo  {journal} {Physical Review B}\ }\textbf {\bibinfo
  {volume} {54}},\ \bibinfo {pages} {1853} (\bibinfo {year}
  {1996})}\BibitemShut {NoStop}%
\bibitem [{\citenamefont {Moessner}\ and\ \citenamefont
  {Chalker}(1996)}]{moessner_exact_1996}%
  \BibitemOpen
  \bibfield  {author} {\bibinfo {author} {\bibfnamefont {R.}~\bibnamefont
  {Moessner}}\ and\ \bibinfo {author} {\bibfnamefont {J.~T.}\ \bibnamefont
  {Chalker}},\ }\href {\doibase 10.1103/PhysRevB.54.5006} {\bibfield  {journal}
  {\bibinfo  {journal} {Physical Review B}\ }\textbf {\bibinfo {volume} {54}},\
  \bibinfo {pages} {5006} (\bibinfo {year} {1996})}\BibitemShut {NoStop}%
\bibitem [{\citenamefont {Shibata}\ and\ \citenamefont
  {Yoshioka}(2001)}]{shibata_ground-state_2001}%
  \BibitemOpen
  \bibfield  {author} {\bibinfo {author} {\bibfnamefont {N.}~\bibnamefont
  {Shibata}}\ and\ \bibinfo {author} {\bibfnamefont {D.}~\bibnamefont
  {Yoshioka}},\ }\href {\doibase 10.1103/PhysRevLett.86.5755} {\bibfield
  {journal} {\bibinfo  {journal} {Physical Review Letters}\ }\textbf {\bibinfo
  {volume} {86}},\ \bibinfo {pages} {5755} (\bibinfo {year}
  {2001})}\BibitemShut {NoStop}%
\bibitem [{\citenamefont {Koulakov}\ \emph {et~al.}(1996)\citenamefont
  {Koulakov}, \citenamefont {Fogler},\ and\ \citenamefont
  {Shklovskii}}]{koulakov_charge_1996}%
  \BibitemOpen
  \bibfield  {author} {\bibinfo {author} {\bibfnamefont {A.~A.}\ \bibnamefont
  {Koulakov}}, \bibinfo {author} {\bibfnamefont {M.~M.}\ \bibnamefont
  {Fogler}}, \ and\ \bibinfo {author} {\bibfnamefont {B.~I.}\ \bibnamefont
  {Shklovskii}},\ }\href {\doibase 10.1103/PhysRevLett.76.499} {\bibfield
  {journal} {\bibinfo  {journal} {Physical Review Letters}\ }\textbf {\bibinfo
  {volume} {76}},\ \bibinfo {pages} {499} (\bibinfo {year} {1996})}\BibitemShut
  {NoStop}%
\bibitem [{\citenamefont {Fogler}(2001)}]{fogler_stripe_2001}%
  \BibitemOpen
  \bibfield  {author} {\bibinfo {author} {\bibfnamefont {M.~M.}\ \bibnamefont
  {Fogler}},\ }\href {http://arxiv.org/abs/cond-mat/0111001} {\bibfield
  {journal} {\bibinfo  {journal} {{arXiv}:cond-mat/0111001}\ } (\bibinfo {year}
  {2001})},\ \bibinfo {note} {pp. 98-138, in High Magnetic Fields: Applications
  in Condensed Matter Physics and Spectroscopy, ed. by C. Berthier, L.-P. Levy,
  G. Martinez (Springer-Verlag, Berlin, 2002)}\BibitemShut {NoStop}%
\bibitem [{\citenamefont {C{\^o}t{\'e}}\ \emph {et~al.}(2003)\citenamefont
  {C{\^o}t{\'e}}, \citenamefont {Doiron}, \citenamefont {Bourassa},\ and\
  \citenamefont {Fertig}}]{cote_dynamics_2003}%
  \BibitemOpen
  \bibfield  {author} {\bibinfo {author} {\bibfnamefont {R.}~\bibnamefont
  {C{\^o}t{\'e}}}, \bibinfo {author} {\bibfnamefont {C.~B.}\ \bibnamefont
  {Doiron}}, \bibinfo {author} {\bibfnamefont {J.}~\bibnamefont {Bourassa}}, \
  and\ \bibinfo {author} {\bibfnamefont {H.~A.}\ \bibnamefont {Fertig}},\
  }\href {\doibase 10.1103/PhysRevB.68.155327} {\bibfield  {journal} {\bibinfo
  {journal} {Physical Review B}\ }\textbf {\bibinfo {volume} {68}},\ \bibinfo
  {pages} {155327} (\bibinfo {year} {2003})}\BibitemShut {NoStop}%
\bibitem [{\citenamefont {Lewis}\ \emph {et~al.}(2002)\citenamefont {Lewis},
  \citenamefont {Ye}, \citenamefont {Engel}, \citenamefont {Tsui},
  \citenamefont {Pfeiffer},\ and\ \citenamefont {West}}]{lewis_microwave_2002}%
  \BibitemOpen
  \bibfield  {author} {\bibinfo {author} {\bibfnamefont {R.~M.}\ \bibnamefont
  {Lewis}}, \bibinfo {author} {\bibfnamefont {P.~D.}\ \bibnamefont {Ye}},
  \bibinfo {author} {\bibfnamefont {L.~W.}\ \bibnamefont {Engel}}, \bibinfo
  {author} {\bibfnamefont {D.~C.}\ \bibnamefont {Tsui}}, \bibinfo {author}
  {\bibfnamefont {L.~N.}\ \bibnamefont {Pfeiffer}}, \ and\ \bibinfo {author}
  {\bibfnamefont {K.~W.}\ \bibnamefont {West}},\ }\href {\doibase
  10.1103/PhysRevLett.89.136804} {\bibfield  {journal} {\bibinfo  {journal}
  {Physical Review Letters}\ }\textbf {\bibinfo {volume} {89}},\ \bibinfo
  {pages} {136804} (\bibinfo {year} {2002})}\BibitemShut {NoStop}%
\bibitem [{\citenamefont {Lewis}\ \emph {et~al.}(2004)\citenamefont {Lewis},
  \citenamefont {Chen}, \citenamefont {Engel}, \citenamefont {Tsui},
  \citenamefont {Ye}, \citenamefont {Pfeiffer},\ and\ \citenamefont
  {West}}]{lewis_evidence_2004}%
  \BibitemOpen
  \bibfield  {author} {\bibinfo {author} {\bibfnamefont {R.~M.}\ \bibnamefont
  {Lewis}}, \bibinfo {author} {\bibfnamefont {Y.}~\bibnamefont {Chen}},
  \bibinfo {author} {\bibfnamefont {L.~W.}\ \bibnamefont {Engel}}, \bibinfo
  {author} {\bibfnamefont {D.~C.}\ \bibnamefont {Tsui}}, \bibinfo {author}
  {\bibfnamefont {P.~D.}\ \bibnamefont {Ye}}, \bibinfo {author} {\bibfnamefont
  {L.~N.}\ \bibnamefont {Pfeiffer}}, \ and\ \bibinfo {author} {\bibfnamefont
  {K.~W.}\ \bibnamefont {West}},\ }\href {\doibase
  10.1103/PhysRevLett.93.176808} {\bibfield  {journal} {\bibinfo  {journal}
  {Physical Review Letters}\ }\textbf {\bibinfo {volume} {93}},\ \bibinfo
  {pages} {176808} (\bibinfo {year} {2004})}\BibitemShut {NoStop}%
\bibitem [{\citenamefont {Lewis}\ \emph {et~al.}(2005)\citenamefont {Lewis},
  \citenamefont {Chen}, \citenamefont {Engel}, \citenamefont {Tsui},
  \citenamefont {Pfeiffer},\ and\ \citenamefont {West}}]{lewis_microwave_2005}%
  \BibitemOpen
  \bibfield  {author} {\bibinfo {author} {\bibfnamefont {R.~M.}\ \bibnamefont
  {Lewis}}, \bibinfo {author} {\bibfnamefont {Y.~P.}\ \bibnamefont {Chen}},
  \bibinfo {author} {\bibfnamefont {L.~W.}\ \bibnamefont {Engel}}, \bibinfo
  {author} {\bibfnamefont {D.~C.}\ \bibnamefont {Tsui}}, \bibinfo {author}
  {\bibfnamefont {L.~N.}\ \bibnamefont {Pfeiffer}}, \ and\ \bibinfo {author}
  {\bibfnamefont {K.~W.}\ \bibnamefont {West}},\ }\href {\doibase
  10.1103/PhysRevB.71.081301} {\bibfield  {journal} {\bibinfo  {journal}
  {Physical Review B}\ }\textbf {\bibinfo {volume} {71}},\ \bibinfo {pages}
  {081301} (\bibinfo {year} {2005})}\BibitemShut {NoStop}%
\bibitem [{\citenamefont {Cooper}\ \emph {et~al.}(2003)\citenamefont {Cooper},
  \citenamefont {Eisenstein}, \citenamefont {Pfeiffer},\ and\ \citenamefont
  {West}}]{cooper_observation_2003}%
  \BibitemOpen
  \bibfield  {author} {\bibinfo {author} {\bibfnamefont {K.~B.}\ \bibnamefont
  {Cooper}}, \bibinfo {author} {\bibfnamefont {J.~P.}\ \bibnamefont
  {Eisenstein}}, \bibinfo {author} {\bibfnamefont {L.~N.}\ \bibnamefont
  {Pfeiffer}}, \ and\ \bibinfo {author} {\bibfnamefont {K.~W.}\ \bibnamefont
  {West}},\ }\href {\doibase 10.1103/PhysRevLett.90.226803} {\bibfield
  {journal} {\bibinfo  {journal} {Physical Review Letters}\ }\textbf {\bibinfo
  {volume} {90}},\ \bibinfo {pages} {226803} (\bibinfo {year}
  {2003})}\BibitemShut {NoStop}%
\bibitem [{\citenamefont {Eisenstein}\ \emph {et~al.}(2002)\citenamefont
  {Eisenstein}, \citenamefont {Cooper}, \citenamefont {Pfeiffer},\ and\
  \citenamefont {West}}]{eisenstein_insulating_2002}%
  \BibitemOpen
  \bibfield  {author} {\bibinfo {author} {\bibfnamefont {J.~P.}\ \bibnamefont
  {Eisenstein}}, \bibinfo {author} {\bibfnamefont {K.~B.}\ \bibnamefont
  {Cooper}}, \bibinfo {author} {\bibfnamefont {L.~N.}\ \bibnamefont
  {Pfeiffer}}, \ and\ \bibinfo {author} {\bibfnamefont {K.~W.}\ \bibnamefont
  {West}},\ }\href {\doibase 10.1103/PhysRevLett.88.076801} {\bibfield
  {journal} {\bibinfo  {journal} {Physical Review Letters}\ }\textbf {\bibinfo
  {volume} {88}},\ \bibinfo {pages} {076801} (\bibinfo {year}
  {2002})}\BibitemShut {NoStop}%
\bibitem [{\citenamefont {Xia}\ \emph {et~al.}(2004)\citenamefont {Xia},
  \citenamefont {Pan}, \citenamefont {Vicente}, \citenamefont {Adams},
  \citenamefont {Sullivan}, \citenamefont {Stormer}, \citenamefont {Tsui},
  \citenamefont {Pfeiffer}, \citenamefont {Baldwin},\ and\ \citenamefont
  {West}}]{xia_electron_2004}%
  \BibitemOpen
  \bibfield  {author} {\bibinfo {author} {\bibfnamefont {J.~S.}\ \bibnamefont
  {Xia}}, \bibinfo {author} {\bibfnamefont {W.}~\bibnamefont {Pan}}, \bibinfo
  {author} {\bibfnamefont {C.~L.}\ \bibnamefont {Vicente}}, \bibinfo {author}
  {\bibfnamefont {E.~D.}\ \bibnamefont {Adams}}, \bibinfo {author}
  {\bibfnamefont {N.~S.}\ \bibnamefont {Sullivan}}, \bibinfo {author}
  {\bibfnamefont {H.~L.}\ \bibnamefont {Stormer}}, \bibinfo {author}
  {\bibfnamefont {D.~C.}\ \bibnamefont {Tsui}}, \bibinfo {author}
  {\bibfnamefont {L.~N.}\ \bibnamefont {Pfeiffer}}, \bibinfo {author}
  {\bibfnamefont {K.~W.}\ \bibnamefont {Baldwin}}, \ and\ \bibinfo {author}
  {\bibfnamefont {K.~W.}\ \bibnamefont {West}},\ }\href {\doibase
  10.1103/PhysRevLett.93.176809} {\bibfield  {journal} {\bibinfo  {journal}
  {Physical Review Letters}\ }\textbf {\bibinfo {volume} {93}},\ \bibinfo
  {pages} {176809} (\bibinfo {year} {2004})}\BibitemShut {NoStop}%
\bibitem [{\citenamefont {Kumar}\ \emph {et~al.}(2010)\citenamefont {Kumar},
  \citenamefont {Cs{\'a}thy}, \citenamefont {Manfra}, \citenamefont
  {Pfeiffer},\ and\ \citenamefont {West}}]{kumar_nonconventional_2010}%
  \BibitemOpen
  \bibfield  {author} {\bibinfo {author} {\bibfnamefont {A.}~\bibnamefont
  {Kumar}}, \bibinfo {author} {\bibfnamefont {G.~A.}\ \bibnamefont
  {Cs{\'a}thy}}, \bibinfo {author} {\bibfnamefont {M.~J.}\ \bibnamefont
  {Manfra}}, \bibinfo {author} {\bibfnamefont {L.~N.}\ \bibnamefont
  {Pfeiffer}}, \ and\ \bibinfo {author} {\bibfnamefont {K.~W.}\ \bibnamefont
  {West}},\ }\href {\doibase 10.1103/PhysRevLett.105.246808} {\bibfield
  {journal} {\bibinfo  {journal} {Physical Review Letters}\ }\textbf {\bibinfo
  {volume} {105}},\ \bibinfo {pages} {246808} (\bibinfo {year}
  {2010})}\BibitemShut {NoStop}%
\bibitem [{\citenamefont {Deng}\ \emph
  {et~al.}(2012{\natexlab{a}})\citenamefont {Deng}, \citenamefont {Kumar},
  \citenamefont {Manfra}, \citenamefont {Pfeiffer}, \citenamefont {West},\ and\
  \citenamefont {Cs{\'a}thy}}]{deng_collective_2012}%
  \BibitemOpen
  \bibfield  {author} {\bibinfo {author} {\bibfnamefont {N.}~\bibnamefont
  {Deng}}, \bibinfo {author} {\bibfnamefont {A.}~\bibnamefont {Kumar}},
  \bibinfo {author} {\bibfnamefont {M.~J.}\ \bibnamefont {Manfra}}, \bibinfo
  {author} {\bibfnamefont {L.~N.}\ \bibnamefont {Pfeiffer}}, \bibinfo {author}
  {\bibfnamefont {K.~W.}\ \bibnamefont {West}}, \ and\ \bibinfo {author}
  {\bibfnamefont {G.~A.}\ \bibnamefont {Cs{\'a}thy}},\ }\href {\doibase
  10.1103/PhysRevLett.108.086803} {\bibfield  {journal} {\bibinfo  {journal}
  {Physical Review Letters}\ }\textbf {\bibinfo {volume} {108}},\ \bibinfo
  {pages} {086803} (\bibinfo {year} {2012}{\natexlab{a}})}\BibitemShut
  {NoStop}%
\bibitem [{\citenamefont {Deng}\ \emph {et~al.}(2014)\citenamefont {Deng},
  \citenamefont {Gardner}, \citenamefont {Mondal}, \citenamefont {Kleinbaum},
  \citenamefont {Manfra},\ and\ \citenamefont {Cs{\'a}thy}}]{deng_nu52_2014}%
  \BibitemOpen
  \bibfield  {author} {\bibinfo {author} {\bibfnamefont {N.}~\bibnamefont
  {Deng}}, \bibinfo {author} {\bibfnamefont {G.~C.}\ \bibnamefont {Gardner}},
  \bibinfo {author} {\bibfnamefont {S.}~\bibnamefont {Mondal}}, \bibinfo
  {author} {\bibfnamefont {E.}~\bibnamefont {Kleinbaum}}, \bibinfo {author}
  {\bibfnamefont {M.~J.}\ \bibnamefont {Manfra}}, \ and\ \bibinfo {author}
  {\bibfnamefont {G.~A.}\ \bibnamefont {Cs{\'a}thy}},\ }\href {\doibase
  10.1103/PhysRevLett.112.116804} {\bibfield  {journal} {\bibinfo  {journal}
  {Physical Review Letters}\ }\textbf {\bibinfo {volume} {112}},\ \bibinfo
  {pages} {116804} (\bibinfo {year} {2014})}\BibitemShut {NoStop}%
\bibitem [{\citenamefont {Cs{\'a}thy}\ \emph {et~al.}(2005)\citenamefont
  {Cs{\'a}thy}, \citenamefont {Xia}, \citenamefont {Vicente}, \citenamefont
  {Adams}, \citenamefont {Sullivan}, \citenamefont {Stormer}, \citenamefont
  {Tsui}, \citenamefont {Pfeiffer},\ and\ \citenamefont
  {West}}]{csathy_tilt-induced_2005}%
  \BibitemOpen
  \bibfield  {author} {\bibinfo {author} {\bibfnamefont {G.~A.}\ \bibnamefont
  {Cs{\'a}thy}}, \bibinfo {author} {\bibfnamefont {J.~S.}\ \bibnamefont {Xia}},
  \bibinfo {author} {\bibfnamefont {C.~L.}\ \bibnamefont {Vicente}}, \bibinfo
  {author} {\bibfnamefont {E.~D.}\ \bibnamefont {Adams}}, \bibinfo {author}
  {\bibfnamefont {N.~S.}\ \bibnamefont {Sullivan}}, \bibinfo {author}
  {\bibfnamefont {H.~L.}\ \bibnamefont {Stormer}}, \bibinfo {author}
  {\bibfnamefont {D.~C.}\ \bibnamefont {Tsui}}, \bibinfo {author}
  {\bibfnamefont {L.~N.}\ \bibnamefont {Pfeiffer}}, \ and\ \bibinfo {author}
  {\bibfnamefont {K.~W.}\ \bibnamefont {West}},\ }\href {\doibase
  10.1103/PhysRevLett.94.146801} {\bibfield  {journal} {\bibinfo  {journal}
  {Physical Review Letters}\ }\textbf {\bibinfo {volume} {94}},\ \bibinfo
  {pages} {146801} (\bibinfo {year} {2005})}\BibitemShut {NoStop}%
\bibitem [{\citenamefont {Pan}\ \emph {et~al.}(2008)\citenamefont {Pan},
  \citenamefont {Xia}, \citenamefont {Stormer}, \citenamefont {Tsui},
  \citenamefont {Vicente}, \citenamefont {Adams}, \citenamefont {Sullivan},
  \citenamefont {Pfeiffer}, \citenamefont {Baldwin},\ and\ \citenamefont
  {West}}]{pan_experimental_2008}%
  \BibitemOpen
  \bibfield  {author} {\bibinfo {author} {\bibfnamefont {W.}~\bibnamefont
  {Pan}}, \bibinfo {author} {\bibfnamefont {J.~S.}\ \bibnamefont {Xia}},
  \bibinfo {author} {\bibfnamefont {H.~L.}\ \bibnamefont {Stormer}}, \bibinfo
  {author} {\bibfnamefont {D.~C.}\ \bibnamefont {Tsui}}, \bibinfo {author}
  {\bibfnamefont {C.}~\bibnamefont {Vicente}}, \bibinfo {author} {\bibfnamefont
  {E.~D.}\ \bibnamefont {Adams}}, \bibinfo {author} {\bibfnamefont {N.~S.}\
  \bibnamefont {Sullivan}}, \bibinfo {author} {\bibfnamefont {L.~N.}\
  \bibnamefont {Pfeiffer}}, \bibinfo {author} {\bibfnamefont {K.~W.}\
  \bibnamefont {Baldwin}}, \ and\ \bibinfo {author} {\bibfnamefont {K.~W.}\
  \bibnamefont {West}},\ }\href {\doibase 10.1103/PhysRevB.77.075307}
  {\bibfield  {journal} {\bibinfo  {journal} {Physical Review B}\ }\textbf
  {\bibinfo {volume} {77}},\ \bibinfo {pages} {075307} (\bibinfo {year}
  {2008})}\BibitemShut {NoStop}%
\bibitem [{\citenamefont {Nuebler}\ \emph {et~al.}(2010)\citenamefont
  {Nuebler}, \citenamefont {Umansky}, \citenamefont {Morf}, \citenamefont
  {Heiblum}, \citenamefont {von Klitzing},\ and\ \citenamefont
  {Smet}}]{nuebler_density_2010}%
  \BibitemOpen
  \bibfield  {author} {\bibinfo {author} {\bibfnamefont {J.}~\bibnamefont
  {Nuebler}}, \bibinfo {author} {\bibfnamefont {V.}~\bibnamefont {Umansky}},
  \bibinfo {author} {\bibfnamefont {R.}~\bibnamefont {Morf}}, \bibinfo {author}
  {\bibfnamefont {M.}~\bibnamefont {Heiblum}}, \bibinfo {author} {\bibfnamefont
  {K.}~\bibnamefont {von Klitzing}}, \ and\ \bibinfo {author} {\bibfnamefont
  {J.}~\bibnamefont {Smet}},\ }\href {\doibase 10.1103/PhysRevB.81.035316}
  {\bibfield  {journal} {\bibinfo  {journal} {Physical Review B}\ }\textbf
  {\bibinfo {volume} {81}},\ \bibinfo {pages} {035316} (\bibinfo {year}
  {2010})}\BibitemShut {NoStop}%
\bibitem [{\citenamefont {Goerbig}\ \emph {et~al.}(2003)\citenamefont
  {Goerbig}, \citenamefont {Lederer},\ and\ \citenamefont
  {Morais~Smith}}]{goerbig_microscopic_2003}%
  \BibitemOpen
  \bibfield  {author} {\bibinfo {author} {\bibfnamefont {M.~O.}\ \bibnamefont
  {Goerbig}}, \bibinfo {author} {\bibfnamefont {P.}~\bibnamefont {Lederer}}, \
  and\ \bibinfo {author} {\bibfnamefont {C.}~\bibnamefont {Morais~Smith}},\
  }\href {\doibase 10.1103/PhysRevB.68.241302} {\bibfield  {journal} {\bibinfo
  {journal} {Physical Review B}\ }\textbf {\bibinfo {volume} {68}},\ \bibinfo
  {pages} {241302} (\bibinfo {year} {2003})}\BibitemShut {NoStop}%
\bibitem [{\citenamefont {Goerbig}\ \emph {et~al.}(2004)\citenamefont
  {Goerbig}, \citenamefont {Lederer},\ and\ \citenamefont
  {Smith}}]{goerbig_competition_2004}%
  \BibitemOpen
  \bibfield  {author} {\bibinfo {author} {\bibfnamefont {M.~O.}\ \bibnamefont
  {Goerbig}}, \bibinfo {author} {\bibfnamefont {P.}~\bibnamefont {Lederer}}, \
  and\ \bibinfo {author} {\bibfnamefont {C.~M.}\ \bibnamefont {Smith}},\ }\href
  {\doibase 10.1103/PhysRevB.69.115327} {\bibfield  {journal} {\bibinfo
  {journal} {Physical Review B}\ }\textbf {\bibinfo {volume} {69}},\ \bibinfo
  {pages} {115327} (\bibinfo {year} {2004})}\BibitemShut {NoStop}%
\bibitem [{\citenamefont {Willett}(2013)}]{willett_quantum_2013}%
  \BibitemOpen
  \bibfield  {author} {\bibinfo {author} {\bibfnamefont {R.~L.}\ \bibnamefont
  {Willett}},\ }\href {\doibase 10.1088/0034-4885/76/7/076501} {\bibfield
  {journal} {\bibinfo  {journal} {Reports on Progress in Physics}\ }\textbf
  {\bibinfo {volume} {76}},\ \bibinfo {pages} {076501} (\bibinfo {year}
  {2013})}\BibitemShut {NoStop}%
\bibitem [{\citenamefont {Reichl}\ \emph {et~al.}(2014)\citenamefont {Reichl},
  \citenamefont {Chen}, \citenamefont {Baer}, \citenamefont {R{\"o}ssler},
  \citenamefont {Ihn}, \citenamefont {Ensslin}, \citenamefont {Dietsche},\ and\
  \citenamefont {Wegscheider}}]{reichl_increasing_2014}%
  \BibitemOpen
  \bibfield  {author} {\bibinfo {author} {\bibfnamefont {C.}~\bibnamefont
  {Reichl}}, \bibinfo {author} {\bibfnamefont {J.}~\bibnamefont {Chen}},
  \bibinfo {author} {\bibfnamefont {S.}~\bibnamefont {Baer}}, \bibinfo {author}
  {\bibfnamefont {C.}~\bibnamefont {R{\"o}ssler}}, \bibinfo {author}
  {\bibfnamefont {T.}~\bibnamefont {Ihn}}, \bibinfo {author} {\bibfnamefont
  {K.}~\bibnamefont {Ensslin}}, \bibinfo {author} {\bibfnamefont
  {W.}~\bibnamefont {Dietsche}}, \ and\ \bibinfo {author} {\bibfnamefont
  {W.}~\bibnamefont {Wegscheider}},\ }\href {\doibase
  10.1088/1367-2630/16/2/023014} {\bibfield  {journal} {\bibinfo  {journal}
  {New Journal of Physics}\ }\textbf {\bibinfo {volume} {16}},\ \bibinfo
  {pages} {023014} (\bibinfo {year} {2014})}\BibitemShut {NoStop}%
\bibitem [{\citenamefont {Baer}(2014)}]{baer_transport_2014}%
  \BibitemOpen
  \bibfield  {author} {\bibinfo {author} {\bibfnamefont {S.}~\bibnamefont
  {Baer}},\ }\emph {\bibinfo {title} {Transport spectroscopy of confined
  fractional quantum Hall systems}},\ \href
  {http://dx.doi.org/10.3929/ethz-a-010252603} {Ph.D. thesis},\ \bibinfo
  {school} {{ETH} Z\"urich}, \bibinfo {address} {Z\"urich} (\bibinfo {year}
  {2014})\BibitemShut {NoStop}%
\bibitem [{\citenamefont {G{\"o}res}\ \emph {et~al.}(2007)\citenamefont
  {G{\"o}res}, \citenamefont {Gamez}, \citenamefont {Smet}, \citenamefont
  {Pfeiffer}, \citenamefont {West}, \citenamefont {Yacoby}, \citenamefont
  {Umansky},\ and\ \citenamefont {von Klitzing}}]{gores_current-induced_2007}%
  \BibitemOpen
  \bibfield  {author} {\bibinfo {author} {\bibfnamefont {J.}~\bibnamefont
  {G{\"o}res}}, \bibinfo {author} {\bibfnamefont {G.}~\bibnamefont {Gamez}},
  \bibinfo {author} {\bibfnamefont {J.~H.}\ \bibnamefont {Smet}}, \bibinfo
  {author} {\bibfnamefont {L.}~\bibnamefont {Pfeiffer}}, \bibinfo {author}
  {\bibfnamefont {K.}~\bibnamefont {West}}, \bibinfo {author} {\bibfnamefont
  {A.}~\bibnamefont {Yacoby}}, \bibinfo {author} {\bibfnamefont
  {V.}~\bibnamefont {Umansky}}, \ and\ \bibinfo {author} {\bibfnamefont
  {K.}~\bibnamefont {von Klitzing}},\ }\href {\doibase
  10.1103/PhysRevLett.99.246402} {\bibfield  {journal} {\bibinfo  {journal}
  {Physical Review Letters}\ }\textbf {\bibinfo {volume} {99}},\ \bibinfo
  {pages} {246402} (\bibinfo {year} {2007})}\BibitemShut {NoStop}%
\bibitem [{\citenamefont {Deng}\ \emph
  {et~al.}(2012{\natexlab{b}})\citenamefont {Deng}, \citenamefont {Watson},
  \citenamefont {Rokhinson}, \citenamefont {Manfra},\ and\ \citenamefont
  {Cs{\'a}thy}}]{deng_contrasting_2012}%
  \BibitemOpen
  \bibfield  {author} {\bibinfo {author} {\bibfnamefont {N.}~\bibnamefont
  {Deng}}, \bibinfo {author} {\bibfnamefont {J.~D.}\ \bibnamefont {Watson}},
  \bibinfo {author} {\bibfnamefont {L.~P.}\ \bibnamefont {Rokhinson}}, \bibinfo
  {author} {\bibfnamefont {M.~J.}\ \bibnamefont {Manfra}}, \ and\ \bibinfo
  {author} {\bibfnamefont {G.~A.}\ \bibnamefont {Cs{\'a}thy}},\ }\href
  {\doibase 10.1103/PhysRevB.86.201301} {\bibfield  {journal} {\bibinfo
  {journal} {Physical Review B}\ }\textbf {\bibinfo {volume} {86}},\ \bibinfo
  {pages} {201301} (\bibinfo {year} {2012}{\natexlab{b}})}\BibitemShut
  {NoStop}%
\bibitem [{\citenamefont {Kosterlitz}\ and\ \citenamefont
  {Thouless}(1973)}]{kosterlitz_ordering_1973}%
  \BibitemOpen
  \bibfield  {author} {\bibinfo {author} {\bibfnamefont {J.~M.}\ \bibnamefont
  {Kosterlitz}}\ and\ \bibinfo {author} {\bibfnamefont {D.~J.}\ \bibnamefont
  {Thouless}},\ }\href {\doibase 10.1088/0022-3719/6/7/010} {\bibfield
  {journal} {\bibinfo  {journal} {Journal of Physics C: Solid State Physics}\
  }\textbf {\bibinfo {volume} {6}},\ \bibinfo {pages} {1181} (\bibinfo {year}
  {1973})}\BibitemShut {NoStop}%
\bibitem [{\citenamefont {Kosterlitz}(1974)}]{kosterlitz_critical_1974}%
  \BibitemOpen
  \bibfield  {author} {\bibinfo {author} {\bibfnamefont {J.~M.}\ \bibnamefont
  {Kosterlitz}},\ }\href {\doibase 10.1088/0022-3719/7/6/005} {\bibfield
  {journal} {\bibinfo  {journal} {Journal of Physics C: Solid State Physics}\
  }\textbf {\bibinfo {volume} {7}},\ \bibinfo {pages} {1046} (\bibinfo {year}
  {1974})}\BibitemShut {NoStop}%
\bibitem [{\citenamefont {Halperin}\ and\ \citenamefont
  {Nelson}(1978)}]{halperin_theory_1978}%
  \BibitemOpen
  \bibfield  {author} {\bibinfo {author} {\bibfnamefont {B.~I.}\ \bibnamefont
  {Halperin}}\ and\ \bibinfo {author} {\bibfnamefont {D.~R.}\ \bibnamefont
  {Nelson}},\ }\href {\doibase 10.1103/PhysRevLett.41.121} {\bibfield
  {journal} {\bibinfo  {journal} {Physical Review Letters}\ }\textbf {\bibinfo
  {volume} {41}},\ \bibinfo {pages} {121} (\bibinfo {year} {1978})}\BibitemShut
  {NoStop}%
\bibitem [{\citenamefont {Young}(1979)}]{young_melting_1979}%
  \BibitemOpen
  \bibfield  {author} {\bibinfo {author} {\bibfnamefont {A.~P.}\ \bibnamefont
  {Young}},\ }\href {\doibase 10.1103/PhysRevB.19.1855} {\bibfield  {journal}
  {\bibinfo  {journal} {Physical Review B}\ }\textbf {\bibinfo {volume} {19}},\
  \bibinfo {pages} {1855} (\bibinfo {year} {1979})}\BibitemShut {NoStop}%
\bibitem [{\citenamefont {Fradkin}\ and\ \citenamefont
  {Kivelson}(1999)}]{fradkin_liquid-crystal_1999}%
  \BibitemOpen
  \bibfield  {author} {\bibinfo {author} {\bibfnamefont {E.}~\bibnamefont
  {Fradkin}}\ and\ \bibinfo {author} {\bibfnamefont {S.~A.}\ \bibnamefont
  {Kivelson}},\ }\href {\doibase 10.1103/PhysRevB.59.8065} {\bibfield
  {journal} {\bibinfo  {journal} {Physical Review B}\ }\textbf {\bibinfo
  {volume} {59}},\ \bibinfo {pages} {8065} (\bibinfo {year}
  {1999})}\BibitemShut {NoStop}%
\bibitem [{\citenamefont {Beenakker}\ and\ \citenamefont {van
  Houten}(2004)}]{beenakker_quantum_2004}%
  \BibitemOpen
  \bibfield  {author} {\bibinfo {author} {\bibfnamefont {C.~W.~J.}\
  \bibnamefont {Beenakker}}\ and\ \bibinfo {author} {\bibfnamefont
  {H.}~\bibnamefont {van Houten}},\ }\href
  {http://arxiv.org/abs/cond-mat/0412664} {\bibfield  {journal} {\bibinfo
  {journal} {cond-mat/0412664}\ } (\bibinfo {year} {2004})},\ \bibinfo {note}
  {\normalfont{Solid State Physics} \textbf{44}, 1 (1991)}\BibitemShut
  {NoStop}%
\bibitem [{\citenamefont {Baer}\ \emph {et~al.}(2014)\citenamefont {Baer},
  \citenamefont {R{\"o}ssler}, \citenamefont {Ihn}, \citenamefont {Ensslin},
  \citenamefont {Reichl},\ and\ \citenamefont
  {Wegscheider}}]{baer_experimental_2014}%
  \BibitemOpen
  \bibfield  {author} {\bibinfo {author} {\bibfnamefont {S.}~\bibnamefont
  {Baer}}, \bibinfo {author} {\bibfnamefont {C.}~\bibnamefont {R{\"o}ssler}},
  \bibinfo {author} {\bibfnamefont {T.}~\bibnamefont {Ihn}}, \bibinfo {author}
  {\bibfnamefont {K.}~\bibnamefont {Ensslin}}, \bibinfo {author} {\bibfnamefont
  {C.}~\bibnamefont {Reichl}}, \ and\ \bibinfo {author} {\bibfnamefont
  {W.}~\bibnamefont {Wegscheider}},\ }\href {\doibase
  10.1103/PhysRevB.90.075403} {\bibfield  {journal} {\bibinfo  {journal}
  {Physical Review B}\ }\textbf {\bibinfo {volume} {90}},\ \bibinfo {pages}
  {075403} (\bibinfo {year} {2014})}\BibitemShut {NoStop}%
\bibitem [{\citenamefont {Rossokhaty}\ \emph {et~al.}(2014)\citenamefont
  {Rossokhaty}, \citenamefont {L\"uscher}, \citenamefont {Folk}, \citenamefont
  {Watson}, \citenamefont {Gardner},\ and\ \citenamefont
  {Manfra}}]{rossokhaty_bias-induced_2014}%
  \BibitemOpen
  \bibfield  {author} {\bibinfo {author} {\bibfnamefont {A.~V.}\ \bibnamefont
  {Rossokhaty}}, \bibinfo {author} {\bibfnamefont {S.}~\bibnamefont
  {L\"uscher}}, \bibinfo {author} {\bibfnamefont {J.~A.}\ \bibnamefont {Folk}},
  \bibinfo {author} {\bibfnamefont {J.~D.}\ \bibnamefont {Watson}}, \bibinfo
  {author} {\bibfnamefont {G.~C.}\ \bibnamefont {Gardner}}, \ and\ \bibinfo
  {author} {\bibfnamefont {M.~J.}\ \bibnamefont {Manfra}},\ }\href
  {http://arxiv.org/abs/1412.1921} {\bibfield  {journal} {\bibinfo  {journal}
  {{arXiv}:1412.1921 [cond-mat]}\ } (\bibinfo {year} {2014})},\ \bibinfo {note}
  {{arXiv}: 1412.1921}\BibitemShut {NoStop}%
\end{thebibliography}%
\end{document}